\documentclass[11pt]{article}
\usepackage{cite}
\usepackage{scalerel}
\usepackage{shuffle}
\usepackage{amsmath,amsfonts,amssymb}
\usepackage{latexsym,epsfig}
\usepackage{dsfont}
\usepackage{hyperref}
\usepackage{extarrows}
\usepackage{comment} 
\usepackage{tikz-cd}

\def\hybrid{
        \topmargin -20pt
        \oddsidemargin 0pt
        \headheight 0pt \headsep 0pt
        \textwidth 6.25in 
        \textheight 9.5in 
        \marginparwidth .875in
        \parskip 5pt plus 1pt \jot = 1.5ex}

\hybrid

 \csname
@addtoreset\endcsname{equation}{section}

\def\be{\begin{equation}}
\def\ee{\end{equation}}

\allowdisplaybreaks

\def\bpm{\begin{pmatrix}}
\def\epm{\end{pmatrix}}

\begin{document}

\begin{titlepage}
\rightline{}
\rightline{December 2025}
\rightline{HU-EP-25/39-RTG}  
\begin{center}
\vskip 1.5cm
{\Large \bf{Homotopy transfer  for 
massive Kaluza-Klein modes  }}
\vskip 1.2cm

{\large\bf {Camille Eloy$^1$, Olaf Hohm$^2$, Camilla Lavino$^2$, \\[1ex]  Henning Samtleben$^{1,3}$ and  Yehudi Simon$^1$ }} \vskip 1.1cm

$^1$ {\it  ENS de Lyon, CNRS, LPENSL, UMR5672, 69342, Lyon cedex 07, France}\\[2.5ex]

$^2$ {\it  Institute for Physics, Humboldt University Berlin,\\
 Zum Gro\ss en Windkanal 6, D-12489 Berlin, Germany}\\[2.5ex]

$^3$ {\it  Institut Universitaire de France (IUF)}\\ [4ex]

 camille.eloy@ens-lyon.fr, ohohm@physik.hu-berlin.de, lavinoca@physik.hu-berlin.de, henning.samtleben@ens-lyon.fr, yehudi.simon@ens-lyon.fr

\vskip .1cm

\vskip .2cm

\end{center}

\bigskip\bigskip
\begin{center} 
\textbf{Abstract}

\end{center} 
\begin{quote}

We develop techniques to treat massive Kaluza-Klein modes to arbitrary  order in perturbation theory. 
The Higgs mechanism that renders the higher Kaluza-Klein modes massive is displayed. To this end we give  
an algorithm in perturbation theory that yields new  fields with the following characteristics: they are gauge invariant under all higher-mode gauge transformations, which  are broken, but they transform covariantly under the zero-mode gauge transformations, 
which are unbroken. 
We employ the formulation of field theory in terms of $L_{\infty}$ algebras together with 
their \textit{homotopy transfer}, which here maps the gauge redundant fields of gravity to gauge invariant fields. 
We illustrate these results, as a proof of concept, for Kaluza-Klein theory on a torus. 
In an accompanying paper these results will be applied to a large class of generalized Scherk-Schwarz backgrounds 
in exceptional field theory.

\end{quote} 
\vfill
\setcounter{footnote}{0}
\end{titlepage}

\tableofcontents

\section{Introduction}

Kaluza-Klein theory denotes a theory in higher dimensions in which some dimensions are compact, say with the background geometry being 
 a torus or a sphere or a more complicated space. The higher-dimensional fields may then be expanded into suitable harmonics of 
this compact space, leading  to a formulation of the 
full higher-dimensional theory that looks like a lower-dimensional theory along the remaining non-compact dimensions, 
but coupled to an infinite tower of  massive fields associated to the higher harmonics. 
These are the massive Kaluza-Klein modes. 
While in phenomenological investigations of extra dimensions one typically aims to determine an `effective' four-dimensional 
theory in which the massive Kaluza-Klein modes are truncated or integrated out, there are contexts 
in which they play an important role and should be kept. The key example for us is holography or the AdS/CFT 
correspondence. For instance, the AdS/CFT duality between type IIB supergravity on $AdS_5\times S^5$ and ${\cal N}=4$ 
super-Yang-Mills theory in four dimensions relates the massive Kaluza-Klein modes corresponding to $S^5$ 
to certain operators on the CFT side.  
More precisely, the Kaluza-Klein spectrum (the free theory) determines the conformal dimensions of these operators, while their higher 
interaction vertices determine (via  Witten diagrams) their correlation functions \cite{Maldacena:1997re,Gubser:1998bc,Witten:1998qj}.

Our goal in this paper, which is the first of a series, is to develop techniques to treat the massive Kaluza-Klein modes 
to arbitrary order in perturbation theory, using homotopy algebra methods  such as homotopy transfer. 
In particular, we will exhibit the Higgs mechanism that renders the 
higher Kaluza-Klein modes massive for the vector, spin-2 and other tensor fields. While it is conceptually 
straightforward to expand a gravity action to any order in fluctuations, 
the resulting interaction terms are a mixture of the physical Kaluza-Klein modes  and the pure gauge modes, 
thereby  obscuring their physical meaning. One needs to reorganize the fields into the gauge invariant and 
hence physical massive  Kaluza-Klein modes that, thereby, absorb the pure gauge St\"uckelberg fields 
via the Higgs mechanism.

The physical Kaluza-Klein spectrum of  five-dimensional gravity compactified on a circle 
was determined by Salam and Strathdee by analyzing the quadratic theory in a certain gauge 
 \cite{Salam:1981xd}. Subsequently, a more systematic understanding of the underlying Higgs mechanism emerged, 
thanks to the observation  by Duff and Dolan that the higher-dimensional diffeomorphisms imply  an infinite-dimensional symmetry 
that is spontaneously broken by the ground state (the Kaluza-Klein vacuum) \cite{Dolan:1983aa}. 
It was then shown that the fields of the linearized theory can be reorganized into combinations 
that are gauge invariant under the higher mode gauge transformations that 
are spontaneously broken  \cite{Aulakh:1985un,Cho:1992rq}.

However, even for a flat  torus background $T^d$,  it was not until 1999 that the Kaluza-Klein spectrum for pure gravity  
was directly determined by a field theory analysis of the Higgs mechanism, 
which turns out to be surprisingly subtle even at the level of the  free theory \cite{Han:1998sg,Giudice:1998ck}. 
Since then, the Kaluza-Klein spectra on significantly more involved backgrounds including AdS factors 
have become the center of attention, due  to their importance for the AdS/CFT duality, building on the early work of \cite{Sezgin:1983ik,Biran:1983iy,Gunaydin:1984fk,Kim:1985ez}. 
In the recent five years, powerful new techniques were  developed using the  framework  
of exceptional field theory \cite{Hohm:2013pua,Hohm:2013vpa,Hohm:2014qga,Baguet:2015xha}, 
which allow one to determine the Kaluza-Klein spectra of backgrounds with 
little to no symmetries \cite{Malek:2019eaz,Malek:2020yue,Duboeuf:2023cth}. 
However, in this `Kaluza-Klein spectrometry' so far the underlying Higgs mechanisms, particularly for the spin-2 fields,  have not been studied. Rather, the resulting spectra could 
be inferred from the naive mass terms together with an accounting of the Goldstone modes.  
While this procedure was sufficient in order to determine large classes of  mass spectra, it is  
desirable to exhibit explicitly the Higgs mechanism together with  the corresponding rearrangement and diagonalization of the  
fields. This will be particularly important in applications to the AdS/CFT correspondence, 
which relates gauge invariant fields to gauge invariant 
correlation functions, and one would like to write the higher-order couplings in terms 
of gauge invariant fields \cite{Skenderis:2006uy}.

In this paper we will show, as a proof of concept, that the framework of homotopy algebras is ideally 
suited to achieve this goal. While here we treat explicitly  only the toy model of the torus, 
we develop the techniques in all generality so that they will  be applicable to a large class of compactifications, 
including type IIB supergravity on $AdS_5\times S^5$, that  will be analyzed  in an accompanying paper. 
Homotopy algebras denote generalizations of familiar algebras such as Lie algebras, whose homotopy 
versions are known as $L_{\infty}$ algebras \cite{Zwiebach:1992ie}. These are particularly useful since any (semi-classical, perturbative) 
field theory can be encoded in a (cyclic) $L_{\infty}$ algebra \cite{Zeitlin:2007vv,Hohm:2017pnh}, see \cite{Jurco:2018sby} for a review.  
The $L_{\infty}$ formulation  expresses the action of a set of fields, collectively 
denoted by $\Phi$, in terms of multilinear maps or brackets $b_n$ with $n$ arguments  
and an inner product $\langle\,,\rangle$ as 
 \be\label{ActionINTRO} 
  {S}[{\Phi}]
  =\frac{1}{2}\langle {\Phi}, 
  {b}_1({\Phi})\rangle 
  +\frac{1}{3!} \langle {\Phi}, 
  {b}_2({\Phi},{\Phi})\rangle 
  +\frac{1}{4!} \langle {\Phi}, 
{b}_3({\Phi},{\Phi},{\Phi})\rangle +\cdots \;. 
 \ee
For instance, for general relativity, the maps $b_n$ are simply  obtained 
by expanding the Einstein-Hilbert action about a  background spacetime to the desired order. 
However, the $n$-point couplings encoded in $b_{n-1}$ will then consist of a mixture 
of unphysical Goldstone or pure gauge  modes  and the physical modes. 
In order to determine the $n$-point couplings for the physical massive Kaluza-Klein modes 
one can pass over  to field variables  $\widehat{\Phi}$ that are invariant under the non-zero modes of 
the higher-dimensional diffeomorphisms, which are hence spontaneously broken, while the $\widehat{\Phi}$
transform covariantly under the lower-dimensional diffeomorphisms. 
Using the techniques of homotopy transfer\footnote{See \cite{Crainic:2004bxw,Erbin:2020eyc,Koyama:2020qfb,Arvanitakis:2020rrk,Arvanitakis:2021ecw,Bonezzi:2025bgv} for introductions and applications of homotopy transfer.} and the perturbation lemma and employing the results of 
\cite{Chiaffrino:2020akd,Chiaffrino:2023wxk},  we will give an algorithm to compute 
the gauge invariant field variables  to any  order in perturbation theory: 
 \be\label{widehatPhiIntro} 
  \widehat{\Phi} = p_1(\Phi)  + \frac{1}{2} p_2(\Phi, \Phi)  +\frac{1}{3!} p_3(\Phi, \Phi,\Phi) + \cdots\;. 
 \ee
which  generalizes  
the first order and second order results of \cite{Skenderis:2006uy}.

One would expect that  perturbatively rewriting the action (\ref{ActionINTRO}) in terms 
of the gauge invariant combination $\widehat{\Phi}$ leads  to a significant reorganization 
of the action. However, we will prove a result of almost disappointing simplicity: the naive action obtained 
by inserting the gauge covariant (\ref{widehatPhiIntro}) into the original action (\ref{ActionINTRO}), 
 \be\label{SKK} 
  S_{\rm KK}\big[\widehat{\Phi}\big] : = S\big[\widehat{\Phi}\big]\;, 
 \ee
is fully equivalent to (\ref{ActionINTRO}) in the sense that the field equations of (\ref{ActionINTRO}) 
are obeyed if and only if the field equations of (\ref{SKK}) are obeyed. 
Importantly, however, the $\widehat{\Phi}$ is subject to constraints, so that certain couplings of 
the original action become zero in the final action  $S_{\rm KK}$.

The remainder of this paper is organized as follows. In sec.~2 we consider the free (quadratic) part of $D$-dimensional 
gravity on the flat background $M_D = \mathbb{R}^{n-1,1} \times T^d$. We determine the complete quadratic action including 
all Kaluza-Klein modes, and we give a  homotopy transfer interpretation of the gauge invariant fields. 
This will be used to (re-)compute the Kaluza-Klein spectrum in a completely systematic manner. 
In sec.~3 we develop the general theory that allows one to define gauge invariant field variables 
to any order in perturbation theory. In this we make significant use of \cite{Chiaffrino:2020akd} whose results we also simplify 
and generalize. We close with a brief conclusion and outlook section, where we outline the larger research program 
of which this paper is the first step. In an appendix we illustrate homotopy transfer with the toy model 
of massive spin-1 in a St\"uckelberg formulation.

\section{Kaluza-Klein theory on a torus}

In this section we determine the quadratic approximation of  Einstein gravity on the toroidal  Kaluza-Klein background 
$ \mathbb{R}^{n-1,1} \times T^d$, including all massive Kaluza-Klein modes. 
In the first subsection we give this linearized theory in terms of gauge invariant fields for the massive modes, 
thereby exhibiting the Higgs mechanism that renders  spin-2 fields massive. The passing over to gauge invariant 
fields will be interpreted in the second subsection as homotopy transfer, which in turn will be used in the third 
subsection to (re-)derive the Kaluza-Klein spectrum. 

\subsection{Linearized gravity on a $d$-torus }

We begin with the Einstein-Hilbert action in $D$ dimensions for the metric tensor field $G_{MN}$ depending 
on $D$ coordinates $X^M$, $M,N=0,\ldots, D-1$,  
 \be\label{EinsteinHilbert} 
  S = \int d^DX\sqrt{G}\,R(G) \;. 
 \ee 
This theory is invariant under $D$-dimensional diffeomorphisms, which are  infinitesimally generated by the Lie derivative: 
 \be\label{GaugeFull} 
  \delta_{\xi} G_{MN}  = {\cal L}_{\xi}G_{MN} \equiv  \xi^{K}\partial_K G_{MN}  +\partial_M\xi^{K} G_{KN} +\partial_N\xi^{K} G_{MK}\;.  
 \ee 
 
We now expand this theory around a flat background spacetime of the form 
 \be\label{Backgroundspace} 
  M_D = \mathbb{R}^{n-1,1} \times T^d\;, \qquad D=n+d\;, 
 \ee
where $\mathbb{R}^{n-1,1}$ denotes the $n$-dimensional Minkowski space, and $T^d$ is a $d$-torus. 
Thus, we perform the background expansion 
 \be\label{backgroundmetricexp}
  G_{MN}(X)  = \bar{G}_{MN} + h_{MN}(X) \;, \qquad  \bar{G}_{MN}  = \begin{pmatrix} \eta_{\mu\nu} &  0 \\ 0 & \bar{G}_{mn} \end{pmatrix} \;, 
 \ee
where $\bar{G}_{mn}$ denotes the constant metric on the torus (that  we could take to be $\bar{G}_{mn}=\delta_{mn}$). 
To quadratic order in fields, (\ref{EinsteinHilbert}) reduces to  
 \be\label{quadraticEinstein} 
  S = -\frac{1}{2} \int d^{D} X\, h^{MN} {\cal G}^D_{MN} \;, 
 \ee
where 
 \be\label{Einsteinandshit} 
 \begin{split} 
  {\cal G}^D_{MN} &= R^D_{MN} - \frac{1}{2}R^D\,\bar{G}_{MN}\;,  \\
  R_{MN}^{D} &= -\frac{1}{2} \Big(\square_D h_{MN}-2\partial^{K}\partial_{(M}h_{N)K} + \partial_M\partial_N h_D\Big) \;, 
 \end{split} 
 \ee
with  the $D$-dimensional wave operator $\square_D:=\bar{G}^{MN}\partial_M\partial_N$ and the trace 
$h_D:=\bar{G}^{MN}h_{MN}$.  
The action (\ref{quadraticEinstein}) and the tensors (\ref{Einsteinandshit}) are  gauge invariant under 
the transformations following from (\ref{GaugeFull}) to lowest order in fields: 
 \be\label{LinDIff} 
  \delta_{\xi} h_{MN} = \partial_M\xi_N+\partial_N\xi_M \;, 
 \ee
where $\xi_M = \bar{G}_{MN}\xi^{N}$. (More generally, all indices are raised and lowered with the 
flat background metric $\bar{G}_{MN}$.)

 The background spacetime (\ref{Backgroundspace}) warrants a split of fields and indices into the $n$ non-compact or external 
 directions and the $d$ compact or internal  directions. We write for coordinates and indices 
  \be
    X^{M} = (x^{\mu}, y^m)\;, \qquad \mu,\nu = 0,\ldots, n-1\;, \quad m,n=1,\ldots, d\;. 
  \ee
The Kaluza-Klein ansatz then reads 
 \be 
 \begin{split} 
  h_{MN} &= \begin{pmatrix} h_{\mu\nu} & h_{\mu n} \\ h_{m\nu} & h_{mn} \end{pmatrix} 
  = \begin{pmatrix} h_{\mu\nu} & a_{\mu n} \\ a_{\nu m} & \varphi_{mn}  \end{pmatrix} \;. 
 \end{split} 
 \ee
The $D$-dimensional wave operator  decomposes as 
 \be
  \square_D= \square +\Delta \;, \qquad \square = \eta^{\mu\nu}\partial_{\mu}\partial_{\nu}\;, \qquad
  \Delta = \bar{G}^{mn}\partial_m\partial_n\;, 
 \ee
 while the trace decomposes as 
  \be
  h_D = h+\varphi\;, \qquad h: = \eta^{\mu\nu} h_{\mu\nu} \;, \qquad \varphi:=\bar{G}^{mn} \varphi_{mn} \;. 
 \ee
Furthermore, the linearized diffeomorphisms  (\ref{LinDIff}) split as  $\xi^{M} = (\xi^{\mu}, \xi^{m})\equiv (\xi^{\mu}, \lambda^m)$, 
where 
 \be\label{KKgauge} 
 \begin{split} 
  \delta h_{\mu\nu} &= \partial_{\mu}\xi_{\nu} + \partial_{\nu}\xi_{\mu}\;,  \\
  \delta a_{\mu m} &= \partial_{\mu}\lambda_m + \partial_m\xi_{\mu} \;,  \\
  \delta \varphi_{mn} &= \partial_m\lambda_n + \partial_n\lambda_m\;.  
 \end{split} 
 \ee  
 
 Before decomposing the quadratic  action (\ref{quadraticEinstein}) under this Kaluza-Klein split 
 it is beneficial to inspect  the mode expansions of fields and gauge parameters. 
 This is necessary in order to analyze the rearrangement of the fields  into gauge invariant  massive Kaluza-Klein modes
 and pure gauge or St\"uckelberg modes that facilitate the Higgs mechanism. 
 On a torus, the mode expansion is simply given by the Fourier expansion, which for a generic field $\varphi$ reads 
 \be\label{modeexpansionTorus} 
  \varphi(x,y) = \sum_{\Omega\in (\mathbb{Z}^d)^*} \varphi_{\Omega}(x){\cal Y}_{\Omega}(y) \;, \qquad
  {\cal Y}_{\Omega}(y) := e^{i\Omega_m y^m} \;, 
 \ee
where $\Omega_m$ is a $d$-vector with integer entries in order to comply with the 
periodicity of the torus. (The notation $(\mathbb{Z}^d)^*$ indicates that 
if $y^m$ is taken to live in $\mathbb{Z}^d$ then 
$\Omega_m$ lives in the dual lattice.)  
  Consider  the Laplacian 
   \be
    \Delta= \partial^m\partial_m = \bar{G}^{mn} \partial_m\partial_n\;, 
   \ee
 which acts  on (\ref{modeexpansionTorus}) as 
  \be
   \Delta \varphi(x,y) = - \sum_{\Omega\in (\mathbb{Z}^d)^*} \Omega^2 \varphi_{\Omega}(x){\cal Y}_{\Omega}(y) \;, 
  \ee
 where $\Omega^2=\bar{G}^{mn}\Omega_m\Omega_n$.  We now define an operator $K$ acting only on non-zero modes as follows: 
  \be\label{GGreens} 
   (K \varphi)(x,y)  := - \sum_{\Omega\neq 0} (\Omega^2)^{-1} \varphi_{\Omega}(x){\cal Y}_{\Omega}(y) \;. 
  \ee
This operation is well-defined, because $\Omega\neq 0 \Rightarrow \Omega^2\neq0$.   
For the following analysis, we need the projector to zero modes and the orthogonal projector to non-zero modes:   
 \be\label{zerononzerodecomp} 
  [\varphi] := \varphi_{\Omega=0}\;, \qquad \underline{\varphi} := \varphi - [\varphi] \;. 
 \ee
Note that, by definition (\ref{GGreens}), $K$ gives zero on zero modes: 
 \be\label{Gzeromode} 
  K[\varphi] = 0\;. 
 \ee 
Moreover, $K$  is  the inverse to $\Delta$ on non-zero modes in that 
 \be\label{GDeltaINV} 
  \Delta K = 1-[\,\cdot\,]\;, \qquad 
  K\Delta =  1-[\,\cdot\,]\;, 
 \ee  
which in turn can be expressed  as 
 \be\label{InverseG} 
   \Delta K {\varphi} = K\Delta {\varphi} = \underline{\varphi}\;. 
 \ee  
It must be emphasized that $K$ is perfectly well-defined, in sharp contrast to, say, the inverse of the d'Alembert wave operator $\square$ in Minkowski space. Finally, it is easy to see that under the torus integral $K$ is symmetric in that 
 \be\label{Kissymmetric} 
  \int d^dy\,(K\varphi_1) \varphi_2 =   \int d^dy\,\varphi_1 (K\varphi_2) \;.  
 \ee

\subsubsection*{Gauge invariant variables}
Next, inspired by the methods of  cosmological perturbation theory \cite{Bardeen:1980kt,Mukhanov:1990me}, we can perform a so-called scalar-vector-tensor (SVT)  decomposition 
with respect to the internal indices: 
 \be\label{SVTdecomp} 
  \begin{split}
   a_{\mu m} &= \widetilde{a}_{\mu m}+ \partial_mb_{\mu} \;, 
    \\
   \varphi_{mn} &=  \widetilde{\varphi}_{mn} + 2\partial_{(m}\alpha_{n)} + \partial_m\partial_n\beta 
   \;, \\
   \lambda_m &= \widetilde{\lambda}_m + \partial_m\chi\;, 
  \end{split} 
 \ee
where 
 \be\label{constraints} 
  \partial^m \widetilde{a}_{\mu m}=0\;, \qquad  \partial^m\widetilde{\varphi}_{mn} = 0 \;, \qquad \partial^m\alpha_m=0 \;, \qquad 
   \partial^m\widetilde{\lambda}_m=0\;. 
 \ee 
This is just the tensor version of the decomposition of  a vector into a divergence-free vector plus the gradient of a scalar. 
In fact, the SVT components on the right-hand side of (\ref{SVTdecomp}) can be defined in terms of the original fields so that (\ref{SVTdecomp}) 
becomes an identity. To this end we assume that $b_{\mu}$, $\alpha_m$, $\beta$ and $\chi$, which appear only under internal 
derivatives, carry \textit{only non-zero modes}. (Alternatively, if they had zero modes, there would be additional gauge and gauge for gauge 
redundancies, since 
these zero modes drop out of (\ref{SVTdecomp}).) 
We set 
 \be\label{SVTdecomp234} 
 \begin{split}
  b_{\mu} &:= K\partial^na_{\mu n}\;,  \\
  \alpha_n&:= K \partial^m \varphi_{mn} - K^2 \partial_n \partial^p \partial^q \varphi_{pq} \;, \\
  \beta &:=K^2 \partial^m \partial^n \varphi_{mn}\;,  \\
  \chi &:= K\partial^m \lambda_m\;, 
  \end{split}
 \ee
and define the remaining SVT components in terms of these  so that (\ref{SVTdecomp}) are just identities: 
  \be\label{SVTdecomp233} 
  \begin{split}
   \widetilde{a}_{\mu m} &:= a_{\mu m} -  \partial_mb_{\mu} \;, 
    \\
   \widetilde{\varphi}_{mn} &:= \varphi_{mn} -  2\partial_{(m}\alpha_{n)} - \partial_m\partial_n\beta 
   \;, \\
   \widetilde{\lambda}_m  &:=  \lambda_m - \partial_m\chi\;. 
  \end{split} 
 \ee
The non-trivial statement then is  that the constraints (\ref{constraints}) are obeyed. This is easily verified, for instance, 
 \be
  \partial^m \widetilde{a}_{\mu m} = \partial^ma_{\mu m} -  \Delta b_{\mu} =  \partial^ma_{\mu m} -  \Delta K\partial^na_{\mu n} =0\;, 
 \ee 
using (\ref{InverseG}) in the last step, together with  $\partial^na_{\mu n}$ only carrying non-zero modes.

We can now compute the gauge transformations of the SVT components. For instance, acting on the first in (\ref{SVTdecomp}) 
and using (\ref{KKgauge}) we have 
 \be
  \delta a_{\mu m} = \delta\,\widetilde{a}_{\mu m}  + \partial_m(\delta b_{\mu}) =  \partial_{\mu}{\lambda}_m + \partial_m\xi_{\mu}
  = \partial_{\mu}\widetilde{\lambda}_m + \partial_m(\xi_{\mu}+\partial_{\mu}\chi) \;. 
 \ee
Taking the divergence by acting with $\partial^m$ and using the constraints (\ref{constraints})  (which, of course, also hold on field variations) 
one obtains 
 \be
  \Delta(\delta b_{\mu}) = \Delta(\xi_{\mu}+\partial_{\mu}\chi)\;. 
 \ee
Acting on this with $K$ and using (\ref{GDeltaINV}) one obtains $\delta b_{\mu}=\underline{\xi}_{\mu}+\partial_{\mu}\chi$, 
where we recalled that $b_{\mu}$ and $\chi$ carry no zero modes, in contrast to $\xi_{\mu}$. 
Similar manipulations quickly yield the gauge transformations of all SVT components:
 \be\label{SVTgaugetrans} 
 \begin{split} 
  \delta h_{\mu\nu} &= \partial_{\mu}\xi_{\nu} + \partial_{\nu}\xi_{\mu} \\
  \delta \widetilde{a}_{\mu m} &= \partial_{\mu}\widetilde{\lambda}_m \\
  \delta b_{\mu} &= \partial_{\mu}\chi + \underline{\xi}_{\mu} \;, \\
  \delta \widetilde{\varphi}_{mn} &=0\;, \\
  \delta \alpha_n &= \underline{\widetilde{\lambda}}_n \;, \\
  \delta \beta &= 2\chi\;, 
 \end{split} 
 \ee  
where we included the spin-2 field $h_{\mu\nu}$ for completeness.  

With this decomposition we can now build new field variables that are fully 
gauge invariant under all non-zero mode gauge transformations: 
\be\label{gaugeinvFIELDS} 
  \begin{split}
   \widehat{h}_{\mu\nu} &= h_{\mu\nu} - 2\partial_{(\mu}b_{\nu)} + \partial_{\mu}\partial_{\nu}\beta\;, \\
   \widehat{a}_{\mu m} &= \widetilde{a}_{\mu m} - \partial_{\mu}\alpha_m\;, \\
   \widehat{\varphi}_{mn} &= \widetilde{\varphi}_{mn}\;. 
  \end{split} 
 \ee
Indeed, with (\ref{SVTgaugetrans}) and (\ref{zerononzerodecomp}) one quickly computes 
for their gauge transformations: 
   \be\label{zeromodegauge} 
  \begin{split}
   \delta\, \widehat{h}_{\mu\nu} &= \partial_{\mu}[\xi_{\nu}] + \partial_{\nu}[\xi_{\mu}] \;, \\
   \delta\, \widehat{a}_{\mu m} &= \partial_{\mu}[\lambda_m] \;, \\
   \delta\, \widehat{\varphi}_{mn} &= 0\;. 
  \end{split} 
 \ee
Thus, only the zero modes of the gauge parameters are left as genuine gauge symmetries, 
while the non-zero modes, which we say are spontaneously broken, haven been trivialized. 
Note that the above gauge invariant fields obey the constraints that all internal divergencies 
are zero: 
 \be\label{vectorconstraint} 
  \partial^m  \widehat{a}_{\mu m} \equiv  0 \;, \qquad 
   \partial^m \widehat{\varphi}_{mn}   \equiv 0\;, 
 \ee
while the trace $\widehat{\varphi}\equiv  \widehat{\varphi}^{\,m}{}_m$ is still non-zero. These relations  follow immediately from 
(\ref{constraints}). 

As a quick aside let us count the number of degrees of freedom (d.o.f.) that are either gauge invariant and hence physical 
or pure gauge and hence St\"uckelberg fields. 
Since $\widehat{\varphi}_{mn}$ is subject to the $d$ constraints in (\ref{vectorconstraint}) 
it encodes  $\frac{1}{2}d(d+1) - d = \frac{1}{2}d(d-1)$ independent gauge invariant  components, which is hence 
the number of physical scalar fields. 
Of the  $d$ St\"uckelberg scalars not encoded in $ \widehat{\varphi}_{mn}$, 
 $d-1$ get eaten by the vectors 
 $\widehat{a}_{\mu m}$, of which there are $d-1$ due to (\ref{vectorconstraint}), 
  which in turn become massive carrying one more d.o.f.~each. 
 The remaining one 
 scalar gets eaten by the spin-2 field to become massive. 
 The counting of d.o.f.~works out since a  massive spin-2 field in $n$ dimensions carries 
$ \frac{1}{2}n(n-1)-1$ d.o.f., while  a massless one carries $\frac{1}{2}n(n-3) $, 
so the number of St\"uckelberg fields that get eaten should be 
  \be
   \frac{1}{2}n(n-1)-1 -\frac{1}{2}n(n-3)= n-1\;, 
  \ee
corresponding to the eating of one massless vector with $n-2$ d.o.f.~and one massless scalar.

Finally, there are various useful alternative writings of the gauge invariant field variables (\ref{gaugeinvFIELDS}). 
Inserting (\ref{SVTdecomp234}), (\ref{SVTdecomp233}), they can be written explicitly in terms 
of the original fields as  
 \be\label{physicaltorusfields} 
	\begin{split}
	 \widehat{h}_{\mu\nu} &= h_{\mu\nu} - 2K \partial_{(\mu}\partial \cdot  a_{\nu)} 
  +\partial_{\mu}\partial_{\nu} K^2(\partial\cdot\partial\cdot \varphi )\;,  \\ 
		 \widehat{a}_{\mu m} &= a_{\mu m} - \partial_mK(\partial\cdot a_{\mu}) 
  - \partial_{\mu}K (\partial \cdot \varphi)_{m} + \partial_{\mu}\partial_mK^2(\partial\cdot\partial\cdot \varphi )\;,  \\		
 \widehat{\varphi}_{mn}  &= \varphi_{mn} -2K(\partial_{(m}\partial\cdot \varphi_{n)} ) +\partial_m\partial_nK^2(\partial\cdot\partial\cdot \varphi )\;,   \\
  \end{split}
	\ee
where we use the short-hand notation $\partial\cdot \varphi_n = \partial^m\varphi_{mn}$, $\partial\cdot\partial\cdot \varphi=\partial^m\partial^n \varphi_{mn}$, 
etc., for internal divergencies. 	
Moreover, the original fields can be written as the gauge invariant fields plus field-dependent 
pure gauge terms: 
 \be\label{homooTOPP}
 \begin{split} 
  h_{\mu\nu} &= \widehat{h}_{\mu\nu} + \partial_{\mu}\frak{h}(\Phi)_{\nu} + \partial_{\nu}\frak{h}(\Phi)_{\mu} \;, \\
  a_{\mu m} &= \widehat{a}_{\mu m} + \partial_{\mu}\frak{h}(\Phi)_{m}  +\partial_m\frak{h}(\Phi)_{\mu} \;, \\
  \varphi_{mn} &= \widehat{\varphi}_{mn}  + \partial_{m}\frak{h}(\Phi)_{n} + \partial_{n}\frak{h}(\Phi)_{m}\;, 
 \end{split} 
 \ee
where 
 \be\label{homotopy0Torus2} 
  \frak{h}(\Phi) := \begin{pmatrix} \frak{h}(\Phi)_{\mu} \\ \frak{h}(\Phi)_{m} \end{pmatrix} = 
   \begin{pmatrix} b_{\mu} - \tfrac{1}{2} \partial_{\mu}\beta \\ 
   \alpha_m + \frac{1}{2} \partial_{m}\beta  \end{pmatrix} 
   = 
   \begin{pmatrix} K(\partial\cdot a_{\mu}) - \frac{1}{2} \partial_{\mu}(K^2(\partial\cdot\partial\cdot \varphi))  \\ 
   K(\partial\cdot \varphi_{m}) - \frac{1}{2} \partial_{m}(K^2(\partial\cdot\partial\cdot \varphi))  \end{pmatrix} 
   \;, 
 \ee 
where in the last step we inserted (\ref{SVTdecomp234}).

\subsubsection*{Einstein tensor and action}  

We now return to the computation of the Einstein tensor (\ref{Einsteinandshit}) in terms of 
the above gauge invariant Kaluza-Klein field variables. Indeed, since the Einstein tensor is 
gauge invariant, it is clear that it is  writable entirely in terms of (\ref{gaugeinvFIELDS}). 
One finds for the components of the Ricci tensor: 
 \be
 \begin{split}
  R_{\mu\nu}^{D} & = R_{\mu\nu}(\widehat{h})  -\frac{1}{2}\Delta \widehat{h}_{\mu\nu}  -\frac{1}{2} \partial_{\mu}\partial_{\nu}\widehat{\varphi} \;, \\
  R_{\mu n}^{D} & =  -\frac{1}{2} \partial^{\nu} F_{\nu\mu n}(\widehat{a}) -\frac{1}{2} \Delta \widehat{a}_{\mu n} 
  -\frac{1}{2} \partial_{\mu}\partial_n\widehat{\varphi}  
  +\frac{1}{2}
  \partial_n(\partial^{\nu} \widehat{h}_{\nu\mu} - \partial_{\mu}\widehat{h}) \;, \\
    R^D_{mn}  &=  -\frac{1}{2} \square \widehat{\varphi}_{mn} -\frac{1}{2}\Delta  \widehat{\varphi}_{mn} 
  +\partial^{\nu}\partial_{(m}\widehat{a}_{\nu n)} 
  -\frac{1}{2} \partial_m\partial_n(\widehat{h}+\widehat{\varphi}) \;, 
 \end{split}  
 \ee 
where $R_{\mu\nu}$ denotes the purely $n$-dimensional (linearized) Ricci tensor, without any 
internal derivatives, and 
 \be
  F_{\mu\nu m}(\widehat{a})  = \partial_{\mu} \widehat{a}_{\nu m}-\partial_{\nu} \widehat{a}_{\mu m}\;. 
 \ee
(Recall from (\ref{zeromodegauge}) that the zero modes of the hatted fields still transform under the zero-mode  gauge transformations, 
so all terms with only external derivatives have to organize into the familiar gauge invariant form.) 
From this one finds for the Ricci scalar: 
 \be
 \begin{split} 
  R_D &=   R  -\Delta\widehat{h} - (\square+\Delta)  \widehat{\varphi}\;, 
 \end{split} 
 \ee
and then for the components of the Einstein tensor:
	\be\label{Einsteintoruscomp} 
	\begin{split}
		{\cal G}^{D}_{\mu\nu} & = {\cal G}_{\mu\nu} - \frac{1}{2}\Delta (\widehat{h}_{\mu\nu} -\widehat{h}\eta_{\mu\nu})
		-\frac{1}{2}(\partial_{\mu}\partial_{\nu}\widehat{\varphi} -(\square +\Delta) \widehat{\varphi}\eta_{\mu\nu})\;,   \\
		{\cal G}^D_{\mu n} &=  -\frac{1}{2} \partial^{\nu} F_{\nu\mu n}(\widehat{a}) -\frac{1}{2} \Delta \widehat{a}_{\mu n} 
                   -\frac{1}{2} \partial_{\mu}\partial_n\widehat{\varphi}  +\frac{1}{2} \partial_n(\partial^{\nu} \widehat{h}_{\nu\mu} - \partial_{\mu}\widehat{h})\;, \\		
		{\cal G}^{D}_{mn} &= -\frac{1}{2}\partial_m\partial_n\widehat{h} -\frac{1}{2}(R - \Delta\widehat{h})\delta_{mn} 
		-\frac{1}{2}(\square+\Delta) (\widehat{\varphi}_{mn}-\widehat{\varphi}\delta_{mn}) 
		-\frac{1}{2}\partial_m\partial_n\widehat{\varphi} +\partial^{\nu} \partial_{(m}\widehat{a}_{\nu\,n)} \;. 		
	\end{split}
	\ee

Finally, we can insert the Einstein tensor components into (\ref{quadraticEinstein}) in order to determine the 
full (quadratic) action encoding all massive Kaluza-Klein modes: 
 \be \label{ACTIONSTEPPPPPPPP}
  \begin{split} 
  S = -\frac{1}{2} \int d^D X h^{MN} {\cal G}^D_{MN}(h)
  = \int d^D X\Big( -\frac{1}{2}  \widehat{h}^{\mu\nu} {\cal G}^D_{\mu\nu} - \widehat{a}^{\mu m} {\cal G}^D_{\mu m} 
   -\frac{1}{2} \widehat{\varphi}^{mn} {\cal G}^D_{mn}\Big)
  \;. 
 \end{split}  
 \ee
Here we used  that with (\ref{homooTOPP}) the fields can be written 
as the gauge invariant ones, plus pure gauge terms, but the latter drop out from  the gauge invariant  action, 
so that we can immediately replace all fields by their gauge invariant ones. 
Then inserting (\ref{Einsteintoruscomp}) one obtains 
  \be\label{KKquadratic}  
  \begin{split} 
  S    &= \int d^D X\Big(  -\frac{1}{2}  \widehat{h}^{\mu\nu} {\cal G}_{\mu\nu}  
  +\frac{1}{4}  \widehat{h}^{\mu\nu} \Delta \widehat{h}_{\mu\nu} -\frac{1}{4} \widehat{h} \Delta\widehat{h} 
   -\frac{1}{2} \widehat{h} \Delta \widehat{\varphi} +\frac{1}{2} \widehat{\varphi} R \\
  &\qquad\qquad\quad\;   -\frac{1}{4} \widehat{F}^{\mu\nu m}\widehat{F}_{\mu\nu m} +\frac{1}{2} \widehat{a}^{\mu m} \Delta \widehat{a}_{\mu m}  \\
  &\qquad\qquad\quad\; 
  +\frac{1}{4}  \widehat{\varphi}^{mn}(\square+\Delta) \widehat{\varphi}_{mn}
  -\frac{1}{4}  \widehat{\varphi}(\square+\Delta) \widehat{\varphi}\Big) \;. 
 \end{split}  
 \ee
The action is not diagonal, due to the mixing terms between $\widehat{\varphi}$ and $h$ 
in the first line. We will not attempt to diagonalize the action by laborious field redefinitions but instead determine the spectrum by using the  
homotopy transfer interpretation explained in the next subsection.

\subsection{Homotopy transfer}

We will now explain that the passing over from the original fields with infinite-dimensional gauge redundancies 
to the gauge covariant  fields with finite-dimensional gauge redundancy can be interpreted as homotopy transfer. 
One begins by organizing the data of the free theory in terms of a \textit{chain complex}: a sequence of vector spaces (a graded vector space) 
with a linear map (differential) $\partial$ between them. For the case at hand we have a so-called  
4-term chain complex 
\be\label{KKComplex} 
  0 \stackrel{}{\longrightarrow}  X_{-1} \stackrel{{\partial_{-1}}}{\longrightarrow} X_0\stackrel{\partial_0}{\longrightarrow} X_1\stackrel{\partial_1}{\longrightarrow} 
  X_2 \stackrel{}{\longrightarrow}  0 \;, 
 \ee  
where  the subindex denotes the degree of each space, and the differential acts as $\partial_i: X_i \rightarrow X_{i+1}$, 
obeying $\partial_{i+1}\circ \partial_i=0$ or $\partial^2=0$ for short. 
Informally, we think of $X_{-1}$ as the space of gauge parameters,  $X_0$ as the space of fields, $X_1$ as the space of equations of motion, 
and $X_2$ as the space of Noether or Bianchi identities.  Concretely, these are infinite-dimensional vector spaces of certain tensor fields 
on $\mathbb{R}^{n-1,1} \times T^d$, which in the following we indicate by their typical elements: 
 \be\label{XSpaces} 
 \begin{split} 
  X_{-1}& = \big\{ \Lambda= (\xi_{\mu},\lambda_m)\big\} \;, \qquad\qquad  \quad \;\;\;  X_0 = \big\{ \Phi=(h_{\mu\nu}, a_{\mu m},\varphi_{mn})\big\} \;, \\
  X_1 &= \big\{ {\cal E}=(E^{\mu\nu}, E^{\mu m}, E^{mn})\big\}\;,  \qquad X_2 = \big\{ {\cal N}=(N^{\mu}, N^m) \big\} \;. 
 \end{split} 
 \ee
In particular, as vector spaces,  $X_{-1}$ and $X_2$ are isomorphic, as are $X_0$ and $X_1$. 

Next we have to define the differential maps so that $\partial^2=0$. 
These are defined as 
 \be\label{b1upstairsss} 
 \begin{split}
  \partial_{-1}(\Lambda) &= \begin{pmatrix} \partial_{\mu}\xi_{\nu} + \partial_{\nu}\xi_{\mu}  \\
   \partial_{\mu}\lambda_m + \partial_m\xi_{\mu}  \\
    \partial_m\lambda_n + \partial_n\lambda_m \end{pmatrix} ,  \quad \\
  \partial_{0}(\Phi) &= \begin{pmatrix} {\cal G}_{\mu\nu}^D \\ {\cal G}_{\mu m}^{D} \\ {\cal G}_{mn}^{D}\end{pmatrix} , \quad \\
  \partial_{1}({\cal E}) &= \begin{pmatrix} \partial_{\nu} { E}^{\nu\mu} + \partial_{n}{ E}^{\mu n }  \\
  \partial_{\mu}{ E}^{\mu n} + \partial_m{ E}^{mn} \end{pmatrix} , 
   \end{split}
 \ee
where the Einstein tensor components are given by (\ref{Einsteintoruscomp}). 
These encode the linearized gauge transformations as $\delta_{\Lambda}\Phi = \partial_{-1}(\Lambda)$ 
and the equations of motion as $\partial_0(\Phi)=0$. The condition $\partial^2=0$ then amounts to $\partial_{0}\circ \partial_{-1}=0$, 
which encodes gauge invariance of the equations of motion, and $\partial_1\circ \partial_0=0$, which encodes the 
Noether  identities (that originate from the Kaluza-Klein split of the Bianchi identity $\partial^M{\cal G}^D_{MN}=0$).  

In order to write an action such as (\ref{ActionINTRO}) given in the introduction we need an inner product or cyclic 
structure. This is an antisymmetric pairing $\langle \,, \rangle :X\times X\rightarrow \mathbb{R}$ of intrinsic degree $-1$, 
meaning it is only non-zero if the degrees of its arguments sum  to $+1$, satisfying 
 \be\label{cyclicity} 
  \langle x_1, \partial(x_2)\rangle = 
   (-1)^{x_1x_2} \langle x_2, \partial(x_1)\rangle\quad \Leftrightarrow \quad  \langle \partial (x_1), x_2\rangle = 
  - (-1)^{x_1} \langle x_1, \partial(x_2)\rangle\;. 
 \ee
For the given 4-term complex the only non-zero pairings are between degree zero and one and between degrees $-1$ 
and $2$, given by 
 \be\label{KKcyclic} 
 \begin{split} 
  \langle \Phi, {\cal E}\rangle &:= \int d^nx d^dy \Big(-\frac{1}{2} h_{\mu\nu}E^{\mu\nu} - a_{\mu m} E^{\mu m} 
  -\frac{1}{2} \varphi_{mn} E^{mn}\Big) \;, \\
    \langle \Lambda, {\cal N}\rangle &:= \int d^nx d^dy \Big(\xi_{\mu} N^{\mu} +\lambda_m N^m \Big)\;. 
 \end{split}  
 \ee
Here the normalization in the first line has been chosen so that with the differential (\ref{b1upstairsss}) 
the action is reproduced as in (\ref{ACTIONSTEPPPPPPPP}). The condition (\ref{cyclicity}) is obeyed.

Having defined the 4-term chain complex (\ref{KKComplex}) we can consider its \textit{cohomology}: the quotient space of $\partial$ closed 
vectors modulo $\partial$ exact terms. For instance, in degree zero, the cohomology 
 \be
  H_0(X) := \frac{{\rm ker}\,\partial_0}{{\rm im}\,\partial_{-1}}
 \ee
consists of fields $\Phi$ obeying the field equations $\partial_0(\Phi)=0$, where one considers two fields as 
equivalent if they  differ by a gauge transformation, $\Phi \sim \Phi + \partial_{-1}(\Lambda)$. 
Thus, $H_0$ is the space of solutions to the equations of motion modulo gauge transformations, 
which can be viewed as the phase space. In particular, for the case at hand, $H_0$ encodes the Kaluza-Klein spectrum. 
Importantly, the homotopy transfer to a `smaller' space, to be discussed next, preserves the cohomology and thus 
can be viewed as a simplifying intermediate step in the computation of the Kaluza-Klein spectrum.

We now turn to the  homotopy transfer, which maps the chain complex (\ref{KKComplex}) to another 
chain complex with the same cohomology. This second chain complex, which can be though of as that of gauge invariant fields 
and their equations, etc., we write as 
\be\label{circleDIFF}
 0 \stackrel{}{\longrightarrow}  \mathring{X}_{-1} \stackrel{\mathring{\partial}_{-1} }{\longrightarrow} \mathring{X}_0\stackrel{\mathring{\partial}_0}{\longrightarrow} \mathring{X}_1\stackrel{\mathring{\partial}_1}{\longrightarrow} 
  \mathring{X}_2 \stackrel{}{\longrightarrow}  0\;. 
 \ee  
The individual spaces can be viewed as subspaces of  (\ref{XSpaces}), satisfying in particular the  constraints of 
the gauge invariant field variables: 
 \be\label{constrainedspaces} 
 \begin{split} 
  \mathring{X}_{-1}& = \big\{ \mathring{\Lambda}= (\mathring{\xi}_{\mu},\mathring{\lambda}_m)\,\big|\, 
  \mathring{\Lambda}=[\mathring{\Lambda}] \big\} \;, \\
  \mathring{X}_0 &= \big\{\mathring{\Phi}=(\mathring{h}_{\mu\nu}, \mathring{a}_{\mu m},\mathring{\varphi}_{mn})\,\big|\,\partial^m \mathring{a}_{\mu m}=0=
   \partial^m\mathring{\varphi}_{mn} \big\} \;, \\
  \mathring{X}_1 &= \big\{ \mathring{\cal E}=(\mathring{E}^{\mu\nu}, \mathring{E}^{\mu m}, \mathring{E}^{mn}) \,\big| \,
  \partial_m \mathring{E}^{\mu m} = 0 = \partial_m \mathring{E}^{mn} \big\}\;,  \\
  \mathring{X}_2 &= \big\{\mathring{\cal N}=(\mathring{N}^{\mu}, \mathring{N}^m) \,\big|\, \mathring{\cal N} =[\mathring{\cal N}] \big\} \;. 
 \end{split} 
 \ee 
Thus, the spaces of gauge parameters and Noether identities contain only zero modes, while the spaces 
of fields and field equations are subject to the constraints that all internal divergencies are zero. 
For clarity, we often denote the objects in this  new complex 
with a circle on top, to distinguish them conceptually from the elements  of the original complex.

Next we have to define the differential $\mathring{\partial}$. As the differential maps between spaces 
obeying constraints we have to include  suitable projectors. Defining   
 \be\label{Projectors} 
  P_m{}^{n} := \delta_m{}^{n} - \partial_m\partial^n K\;, 
 \ee
we can project a vector $V_m$ to its divergence-free part as 
 \be
  \widehat{V}_m := P_{m}{}^{n} V_{n}\quad \Rightarrow \quad \partial^m\widehat{V}_m=0 \;, 
 \ee
as follows quickly with (\ref{InverseG}). Similarly, the contraction of (\ref{Projectors}) 
with a partial  derivative $\partial_n$ vanishes, e.g., 
  \be\label{consequenceofprojector} 
   P_m{}^{n} \partial_n \Phi = 0 \;. 
  \ee
Further, we can define a projector mapping a 2-tensor into a divergence-free 2-tensor as 
   \be\label{Projectors2} 
  P_{mn}{}^{kl} = P_m{}^{k} P_{n}{}^{l} \quad \Rightarrow \quad 
  P_{mn}{}^{kl} T_{kl} = T_{mn} - 2 K\partial_{(m}\partial^{k} T_{n)k} +K^2\partial_m\partial_n\partial^k\partial^l T_{kl}\;, 
 \ee
where we displayed the action on a symmetric 2-tensor. 
The  differentials in (\ref{circleDIFF}) can then be written as  
 \be\label{mathringpartial} 
 \begin{split}
  \mathring{\partial}_{-1}(\mathring{\Lambda}) = \begin{pmatrix} {\partial}_{\mu}\mathring{\xi}_{\nu} + \partial_{\nu}\mathring{\xi}_{\mu}  \\
   \partial_{\mu}\mathring{\lambda}_m  \\
    0 \end{pmatrix} ,\quad 
  \mathring{\partial}_{0}(\mathring{\Phi}) = \begin{pmatrix} \mathring{\cal G}_{\mu\nu}^D \\ P_{m}{}^{n} \mathring{\cal G}_{\mu n}^{D} \\ 
 P_{mn}{}^{kl}  \mathring{\cal G}_{kl}^{D}\end{pmatrix}  , \quad 
  \mathring{\partial}_{1}(\mathring{\cal E}) = \begin{pmatrix} \partial_{\nu} [\mathring{ E}^{\nu\mu}]  \\
  \partial_{\mu}[\mathring{ E}^{\mu n}] \end{pmatrix} , 
   \end{split}
 \ee
where the notation  $\mathring{\cal G}_{\mu\nu}^D$  indicates  that 
these are the same expressions for the Einstein tensor components as in (\ref{Einsteintoruscomp}), 
except evaluated on fields in $\mathring{X}_0$ obeying the corresponding constraints.   
Note that both $\mathring{\partial}_{0}$ and $\mathring{\partial}_1$ include suitable projectors (onto divergence-free tensors or zero modes, 
respectively), as required by their target spaces. 
Using these  constraints and (\ref{consequenceofprojector}) 
the differential $\mathring{\partial}_{0}$ simplifies  as follows: 
	\be\label{Einsteintoruscompdownstairs} 
	\begin{split}
	 \mathring{\partial}_{0}(\mathring{\Phi}) &= \begin{pmatrix} {\cal G}_{\mu\nu}(\mathring{h})  - \frac{1}{2}\Delta (\mathring{h}_{\mu\nu} -\mathring{h}\eta_{\mu\nu})
		-\frac{1}{2}(\partial_{\mu}\partial_{\nu}\mathring{\varphi} -(\square +\Delta) \mathring{\varphi}\eta_{\mu\nu}) 
		 \\[0.5ex]  
		 -\frac{1}{2} \partial^{\nu} F_{\nu\mu m}(\mathring{a}) -\frac{1}{2} \Delta \mathring{a}_{\mu m}   \\[0.5ex]  
   -\frac{1}{2}P_{mn} \Big(R(\mathring{h})  - \Delta\mathring{h} -(\square + \Delta) \mathring{\varphi} \Big) -\frac{1}{2}(\square+\Delta) \mathring{\varphi}_{mn} \end{pmatrix} 	. 
	\end{split}
	\ee

We are now ready to introduce the notion of homotopy transfer. This requires  a projection map $p:X\rightarrow \mathring{X}$ and an inclusion map
$\iota:\mathring{X}\rightarrow X$, both of intrinsic degree zero, so that they map between the individual space 
in (\ref{KKComplex}), (\ref{circleDIFF}) as 
 \be
  p_k: \,X_k\,\longrightarrow \, \mathring{X}_k\;, \qquad \iota_k: \,\mathring{X}_k\,\longrightarrow \, {X}_k\;. 
 \ee
In addition, there is a homotopy map $\frak{h} : X\rightarrow X$ of intrinsic degree $-1$, i.e.,  $\frak{h}_k:X_k\rightarrow X_{k-1}$, 
so that  
  \be\label{homotopyREL} 
 \begin{split}
  \iota\circ p &= {\rm id} - \partial\circ \frak{h} - \frak{h}\circ \partial\;, \\
  p\circ \iota &= {\rm id}\;. 
 \end{split} 
 \ee
Indicating the degree explicitly, these relations read 
$\iota_k\circ p_k = {\rm id}_{X_k} -\partial_{k-1}\circ \frak{h}_k-\frak{h}_{k+1}\circ \partial_k$ 
and $p_k\circ \iota_k = {\rm id}_{\mathring{X}_k}$. 
In addition we demand that projector and inclusion are so-called chain maps that commute 
with the differential in that:
 \be\label{chainmaps} 
 \begin{split} 
  p \circ \partial = \mathring{\partial} \circ p\;,  \qquad 
  \partial  \circ \iota = \iota \circ \mathring{\partial} \;, 
 \end{split} 
 \ee
with the first one acting on $X$ and the second one acting on $\mathring{X}$.  
If there are homotopy data $p$, $\iota$ and $\frak{h}$ obeying the above relations 
then $X$ and $\mathring{X}$ are called quasi-isomorphic, which means that even though 
$p$ and $\iota$ are not inverse to each other the cohomologies  of $X$ and $\mathring{X}$ are 
isomorphic.

 \begin{figure}[h]
	\begin{centering}
		\label{fig:homotopy transfer}
		\begin{tikzpicture}
			\node at (1,0) {$0$};
			\draw[->] (1.5,0)--(2.5,0);
			\node at (3,0) {$X_{-1}$}; 
			\draw[->] (3.5,0)--(4.5,0);
			\node at (4,0.3) {$\partial$};
			\node at (5,0) {$X_0$}; 
			\draw[->] (5.5,0)--(6.5,0);
			\node at (6,0.3) {$\partial$};
			\node at (7,0) {$X_1$}; 
			\draw[->] (7.5,0)--(8.5,0);
			\node at (8,0.3) {$\partial$};
			\node at (9,0) {$X_2$}; 
			\draw[->] (9.5,0)--(10.5,0);
			\node at (11,0) {$0$}; 
			\node at (1,-2) {$0$};
			\draw[->] (1.5,-2)--(2.5,-2);
			\node at (3,-2) {$\mathring{X}_{-1}$}; 
			\draw[->] (3.5,-2)--(4.5,-2);
			\node at (4,-1.7) {$\mathring{\partial}$};
			\node at (5,-2) {$\mathring{X}_{0}$}; 
			\draw[->] (5.5,-2)--(6.5,-2);
			\node at (6,-1.7) {$\mathring{\partial}$};
			\node at (7,-2) {$\mathring{X}_{1}$}; 
			\draw[->] (7.5,-2)--(8.5,-2);
			\node at (8,-1.7) {$\mathring{\partial}$};
			\node at (9,-2) {$\mathring{X}_{2}$}; 
            \draw[->] (9.5,-2)--(10.5,-2);
			\node at (11,-2) {$0$}; 
			\draw[->] (2.8,-0.3)--(2.8,-1.7);
			\node at (2.5,-1) {$p$};
			\draw[->] (3.2,-1.7)--(3.2,-0.3);
			\node at (3.5,-1) {$\iota$};
			\draw[->] (4.8,-0.3)--(4.8,-1.7);
			\node at (4.5,-1) {$p$};
			\draw[->] (5.2,-1.7)--(5.2,-0.3);
			\node at (5.5,-1) {$\iota$};
			\draw[->] (6.8,-0.3)--(6.8,-1.7);
			\node at (6.5,-1) {$p$};
			\draw[->] (7.2,-1.7)--(7.2,-0.3);
			\node at (7.5,-1) {$\iota$};
			\draw[->] (8.8,-0.3)--(8.8,-1.7);
			\node at (8.5,-1) {$p$};
			\draw[->] (9.2,-1.7)--(9.2,-0.3);
			\node at (9.5,-1) {$\iota$};
			\draw[->] (5,0.3) to [bend right=45] (3,0.3);
			\node at (4,1) {$\frak{h}_0$};
			\draw[->] (7,0.3) to [bend right=45] (5,0.3);
			\node at (6,1) {$0$};
			\draw[->] (9,0.3) to [bend right=45] (7,0.3);
			\node at (8,1) {$\frak{h}_2$};
		\end{tikzpicture}
		\caption{Homotopy transfer from $X$ to $\mathring{X}$.}
	\end{centering}
\end{figure}

Let us now define the homotopy data for Kaluza-Klein theory on a torus. 
The projector is given by 
 \be
 \begin{split}
  p_{-1}(\Lambda) &=  \begin{pmatrix} [\xi_{\mu}] \\  [\lambda_m] \end{pmatrix} \;,  \qquad \qquad \;\;
  p_0(\Phi) = \begin{pmatrix} \widehat{h}_{\mu\nu} \\ \widehat{a}_{\mu m} \\ \widehat{\varphi}_{mn} \end{pmatrix}\;, 
  \\
  p_1({\cal E}) &=\begin{pmatrix} {E}_{\mu\nu} \\ P_m{}^{n} {E}_{\mu n} \\ P_{mn}{}^{kl}  {E}_{kl}\end{pmatrix}\;,  \qquad 
  p_2({\cal N}) = \begin{pmatrix} [N^{\mu}] \\ [N^m] \end{pmatrix}  \;, 
 \end{split}
 \ee
where the hatted quantities in the first line denote the gauge invariant combinations in (\ref{physicaltorusfields}). 
The inclusion maps are trivial in degrees $-1$, $0$ and $2$, in that they simply view an element of $\mathring{X}$ as 
an element of $X$, i.e.~$\iota_{-1}(\mathring{\Lambda}) = \mathring{\Lambda}$, $\iota_{0}(\mathring{\Phi}) = \mathring{\Phi}$
and $ \iota_2(\mathring{\cal N}) = \mathring{\cal N}$. However, in degree $1$ the inclusion is non-trivial and reads 
  \be
 \begin{split}
   \iota_1(\mathring{\cal E}) = \begin{pmatrix} \mathring{E}^{\mu\nu} \\
  \mathring{E}^{\mu m} -K\partial^m\partial_{\nu}\mathring{E}^{\mu\nu} \\
  \mathring{E}^{mn} -2K\partial^{(m} \partial_{\mu} \mathring{E}^{n)\mu} +K^2 \partial^m\partial^n(\partial_{\mu}\partial_{\nu} \mathring{E}^{\mu\nu})
  \end{pmatrix} \;. 
 \end{split}
 \ee
These definitions are such that  the chain map conditions (\ref{chainmaps}) are obeyed, which are verified  by explicit computation. 
For instance, in degree zero we have 
 \be
 \begin{split}
 &\left( \iota_1\circ \mathring{\partial}_0   - \partial_0\circ \iota_0\right)(\mathring{\Phi}) = 
 \iota_1 \begin{pmatrix} \mathring{\cal G}_{\mu\nu}^D \\ P_{m}{}^{n} \mathring{\cal G}_{\mu n}^{D} \\ 
 P_{mn}{}^{kl}  \mathring{\cal G}_{kl}^{D}\end{pmatrix}
 - \begin{pmatrix} \mathring{\cal G}_{\mu\nu}^D \\ \mathring{\cal G}_{\mu m}^{D} \\ \mathring{\cal G}_{mn}^{D}\end{pmatrix} \\
 & \qquad =  \begin{pmatrix} \mathring{\cal G}_{\mu\nu}^D \\ 
 P_{m}{}^{n} \mathring{\cal G}_{\mu n}^{D}- K\partial_m\partial^{\nu} \mathring{\cal G}_{\nu\mu}^D \\ 
 P_{mn}{}^{kl}  \mathring{\cal G}_{kl}^{D}
 -2K\partial_{(m}\partial^{\mu} P_{n)}{}^{k} \mathring{\cal G}_{\mu k}^D+K^2\partial_m\partial_n(\partial^{\mu}\partial^{\nu}\mathring{\cal G}_{\mu\nu}^D)
 \end{pmatrix} 
 - \begin{pmatrix} \mathring{\cal G}_{\mu\nu}^D \\ \mathring{\cal G}_{\mu m}^{D} \\ \mathring{\cal G}_{mn}^{D}\end{pmatrix}\\
 &\qquad= 
   \begin{pmatrix}0  \\ -K\partial_m\big(\partial^{n}\mathring{\cal G}_{\mu n}^D+\partial^{\nu}\mathring{\cal G}_{\mu\nu}^D\big)\\ 
   -2K\partial_{(m}\big(\partial^{\mu}\mathring{\cal G}_{n)\mu}^D + \partial^k \mathring{\cal G}_{n)k}^D\big) 
   +K^2\partial_m\partial_n\big(\partial^k\partial^l \mathring{\cal G}_{kl}^D+ 2\partial^k\partial^{\mu} \mathring{\cal G}^D_{k \mu}
   +\partial^{\mu}\partial^{\nu}\mathring{\cal G}_{\mu\nu}^D\big) \end{pmatrix}
    \\
   &\qquad = 0\;, 
 \end{split}
 \ee
where we used (\ref{Projectors2}) and the Bianchi identities  $\partial^N{\cal G}^D_{MN} = 0$ in the last step.

Next, we have to define the homotopy maps. In degree zero this simply gives a homotopy interpretation to 
the map (\ref{homotopy0Torus2}) found in the previous subsection. Moreover, one finds 
 \be\label{homotopy0Torus3} 
 \begin{split} 
  \frak{h}_0(\Phi) &= \begin{pmatrix} \frak{h}_0(\Phi)_{\mu} \\ \frak{h}_0(\Phi)_{m} \end{pmatrix} = 
   \begin{pmatrix} K(\partial\cdot a_{\mu}) - \frac{1}{2} \partial_{\mu}(K^2(\partial\cdot\partial\cdot \varphi))  \\ 
   K(\partial\cdot \varphi_{m}) - \frac{1}{2} \partial_{m}(K^2(\partial\cdot\partial\cdot \varphi))  \end{pmatrix} \;, \\ 
   \frak{h}_1({\cal E})  &= 0 \;, \\
   \frak{h}_2({\cal N}) &= \begin{pmatrix} \frak{h}_2({\cal N})^{\mu\nu} \\ \frak{h}_2({\cal N})^{\mu m} \\ \frak{h}_2({\cal N})^{mn} \end{pmatrix} 
  = \begin{pmatrix} 0  \\ \partial^mK N^{\mu}   \\2K\partial^{(m}N^{n)} - K^2 \partial^m\partial^n 
  (\partial_{\mu}N^{\mu}  + \partial_k N^k)  \end{pmatrix} \;, 
 \end{split}   
 \ee
while all $\frak{h} $ in other degrees are trivially zero. Furthermore, 
note  that the homotopy map in degree 1 can be chosen to be zero.  
The homotopy relations (\ref{homotopyREL}) are verified by direct computation. 
For instance, in degree 2 we have 
 \be
 \begin{split} 
  ({\bf 1}-\iota p)({\cal N}) &=  ({\bf 1}-\iota p)\begin{pmatrix} N^{\mu} \\ N^m  \end{pmatrix}  
  = \begin{pmatrix} N^{\mu}-[N^{\mu}] \\ N^m - [N^m] \end{pmatrix} = \Delta K\begin{pmatrix} N^{\mu} \\ N^m  \end{pmatrix}  \;, 
 \end{split} 
 \ee
where we used (\ref{GDeltaINV}). Given $\partial_2=0$ this should be equal to 
 \be
 \begin{split} 
  \partial_1(\frak{h}_2({\cal N})) = \begin{pmatrix} \partial_{\nu}\frak{h}_2({\cal N})^{\nu\mu} +\partial_n \frak{h}_2({\cal N})^{\mu n} \\
  \partial_{\mu}\frak{h}_2({\cal N})^{\mu m} + \partial_n \frak{h}_2({\cal N})^{mn} \end{pmatrix} \;, 
 \end{split} 
 \ee
as follows indeed upon inserting the last equation of (\ref{homotopy0Torus3}).

Let us note that the homotopy data obey the so-called side conditions, which will be convenient below: 
 \be\label{sideconditions} 
   p\frak{h} = \frak{h}\iota = \frak{h}^2 = 0 \;. 
 \ee 
This  quickly follows with the formulas above.

Finally, the cyclic structure (\ref{KKcyclic}) is transported to a cyclic structure on $\mathring{X}$: 
 \be\label{transportedcyclci} 
 \big\langle \mathring{x}_1, \mathring{x}_2 \big\rangle_{\mathring{X}} : =  
 \big\langle \iota(\mathring{x}_1) ,\iota(\mathring{x}_2) \big\rangle \;. 
 \ee
The cyclicity condition on $\mathring{X}$  follows from the cyclicity (\ref{cyclicity}) on $X$ together 
with $\iota$ being a chain map.  
 The homotopy $\frak{h}$ is compatible with the cyclic structure in that 
  \be\label{HCyclcic} 
    \big\langle \frak{h}(x_1), x_2\big\rangle  = (-1)^{x_1}\big\langle x_1 , \frak{h}(x_2)\big\rangle \;. 
  \ee
Since in the present case $\frak{h}_1=0$ the only non-trivial instance of this relation  is 
$ \big\langle \frak{h}_0(\Phi), {\cal N} \big\rangle  = \big\langle \Phi, \frak{h}_2({\cal N}) \big\rangle$, 
which follows with a direct computation using (\ref{Kissymmetric}) and  (\ref{homotopy0Torus3}). 
Using the homotopy relation (\ref{homotopyREL}) and the cyclicity  condition it then follows from  (\ref{HCyclcic}) 
that 
 \be
  \big\langle \iota p(x_1), x_2\big\rangle  = \big\langle x_1, \iota p(x_2)\big\rangle\;. 
 \ee
 This implies with (\ref{transportedcyclci}) 
  \be\label{againstcontrainedaergument} 
   \big\langle \mathring{x},p(y)\big\rangle_{\mathring{X}}=   \big\langle \iota(\mathring{x}),\iota p (y)\big\rangle 
   =   \big\langle \iota p\iota(\mathring{x}), y\big\rangle 
    =   \big\langle \iota (\mathring{x}), y\big\rangle  \;, 
  \ee
where we recalled $p\iota={\rm id}$. This relation says that in the pairing against a constrained  object 
belonging to $\mathring{X}$ the projection is automatic and need not be enforced  explicitly.

\subsection{Kaluza-Klein spectrum}

Having established the homotopy transfer to the `smaller' complex $\mathring{X}$ we can now compute its cohomology, 
which is the same as that of the original complex $X$. In particular, the cohomology in degree zero, the space  of solutions modulo gauge transformations, 
determines the Kaluza-Klein spectrum.   

We recall the field equations (\ref{Einsteintoruscompdownstairs}) and denote 
them here as follows: 
 \be\label{downstairsEOM} 
	\begin{split}
		\begin{pmatrix}
			{E}_{\mu\nu}\\[0.5ex]
			{E}_{\mu m}\\[0.5ex]
			{E}_{mn}\\[0.5ex]
		\end{pmatrix}
 &:= \begin{pmatrix} {\cal G}_{\mu\nu}({h})  - \frac{1}{2}\Delta ({h}_{\mu\nu} -{h}\eta_{\mu\nu})
		-\frac{1}{2}(\partial_{\mu}\partial_{\nu}{\varphi} -(\square +\Delta) {\varphi}\eta_{\mu\nu}) 
		\\[0.5ex]  
		-\frac{1}{2} \partial^{\nu} F_{\nu\mu m}({a}) -\frac{1}{2} \Delta {a}_{\mu m}   \\[0.5ex]  
		  -\frac{1}{2}P_{mn} \Big(R({h})  - \Delta{h} -(\square + \Delta) {\varphi} \Big) -\frac{1}{2}(\square+\Delta) {\varphi}_{mn} \end{pmatrix} = 0  	\; , 
\end{split}
\ee
where for ease of reading we removed the circles on top, with the understanding that for the remainder of this section 
all fields obey the constrains encoded in (\ref{constrainedspaces}). 
We will next determine all subsidiary conditions by systematically taking all divergences and traces of these equations. 
Taking divergence, double divergence and trace of $E_{\mu\nu}$ and using the Bianchi identity $\partial^{\mu}{\cal G}_{\mu\nu}=0$ 
we have
\be\label{subsid1} 
\begin{split}
	\begin{pmatrix}
		\partial^\mu{E}_{\mu\nu}\\[0.5ex]
		\partial^\mu\partial^{\nu} {E}_{\mu\nu}\\[0.5ex]
		\eta^{\mu\nu}{E}_{\mu\nu}
	\end{pmatrix}
	&= -\frac{1}{2}\begin{pmatrix} \Delta \left(\partial^\mu {h}_{\mu\nu} - \partial_\nu {h} - \partial_\nu {\varphi} \right)
		\\[0.5ex]  
		 \Delta \left(R({h}) - \square {\varphi}\right)  \\[0.5ex]  
		(n-2) R({h}) - (n-1) \Delta {h} - (n-1) \square {\varphi} - n \Delta {\varphi} \end{pmatrix} =0 	\,, 
\end{split}
\ee
where we used the explicit form of the Ricci scalar $R$, c.f.~(\ref{Einsteinandshit}). 
Next, taking the divergence of $E_{\mu m}$ we obtain 
\be\label{subsid2} 
\partial^\mu{E}_{\mu m} = -\frac{1}{2}\Delta \partial^\mu {a}_{\mu m} = 0 \, .
\ee
Finally, taking the trace of $E_{mn}$ one obtains 
\be\label{subsid3} 
\delta^{mn}{E}_{mn} = -\frac{d}{2}R([h]) + \frac{d-1}{2} \square [\varphi] -\frac{d-1}{2} (R({\underline{h}}) - \Delta \underline{h}) + \frac{d-2}{2} (\square + \Delta) {\underline{\varphi}} = 0\, .
\ee
Acting now with $K$ on the first two equations in (\ref{subsid1}) and using (\ref{GDeltaINV}) we conclude 
\be\label{firstSUBS} 
\begin{split} 
 \partial^\mu {\underline{h}}_{\mu\nu} &= \partial_\nu ({\underline{h}} +  \underline{\varphi})\, ,  \\
 R(\underline{h}) &= \square \underline{\varphi}\;,  
\end{split} 
\ee
which gives a constraint only on the non-zero modes, on which we focus in the following. 
Using $R(\underline{h}) = \square \underline{\varphi}$ in the last equation of (\ref{subsid1}) and in (\ref{subsid3}), we obtain
\be
\begin{split}
    (\square + \Delta) \underline{\varphi} + (n-1) \Delta (\underline{h} + \underline{\varphi}) = 0 \, ,\\
    (\square + \Delta) \underline{\varphi} - (d-1) \Delta (\underline{h} + \underline{\varphi}) = 0 \, .
\end{split}
\ee
Subtracting the second equation from the first yields
\be
(n + d -2) \Delta (\underline{h} + \underline{\varphi}) = 0 \, ,
\ee
and hence, assuming $n+d>2$,
 \be\label{secondSUBS} 
  \underline{h} + \underline{\varphi} = 0\;.  
 \ee
Thus,  with the first equation in (\ref{firstSUBS}), 
\be\label{thirdSUBS} 
  \partial^\mu \underline{h}_{\mu\nu} = 0\;. 
\ee
Similarly, from the equation (\ref{subsid2}) we deduce 
 \be\label{fourthSUBS} 
   \partial^\mu \underline{a}_{\mu m} =0\;.
 \ee
Using then the constraints (\ref{firstSUBS})--(\ref{fourthSUBS}) inside the original equations (\ref{downstairsEOM}) 
one finds that their projections to non-zero modes  reduce to $\underline{E}_{\mu\nu}= -\frac{1}{2}(\square +\Delta)\underline{h}_{\mu\nu}=0$, 
$\underline{E}_{\mu m} = -\tfrac{1}{2} (\square +\Delta)\underline{a}_{\mu m}=0$
and $\underline{E}_{mn} = -\frac{1}{2} (\square +\Delta)\underline{\varphi}_{mn}=0$. 
Summarizing, the spectrum of the non-zero modes is encoded in the following dynamical equations and subsidiary constraints 
 \be\label{FINALDYNEQ} 
  \begin{split}
   (\square +\Delta)\underline{h}_{\mu\nu} &= 0\;, \qquad\; \,  \partial^\mu \underline{h}_{\mu\nu} = 0\;, \qquad\;\; \underline{h} + \underline{\varphi} = 0\;, \\
   (\square +\Delta)\underline{a}_{\mu m} &=0\;, \qquad\; \partial^{\mu}\underline{a}_{\mu m}=0 \;, \qquad \partial^m \underline{a}_{\mu m}=0 \;, \\ 
   (\square +\Delta)\underline{\varphi}_{mn} &= 0\;, \qquad  \partial^m\underline{\varphi}_{mn}=0\;, 
  \end{split}
 \ee
where we also recalled the internal divergence constraints.  Inserting the explicit mode expansion (\ref{modeexpansionTorus}) 
one infers that, for instance, each mode of the spin-2 field obeys the dynamical equations 
 \be
  (\square -M_{\Omega}^2)\underline{h}_{\mu\nu,\Omega}=0\;, 
 \ee
where $M_{\Omega}^2 = \Omega_k\Omega^k$, and similarly for the vectors and scalars. 
One may quickly verify that we have the right number of subsidiary conditions so that 
the total number of propagating degrees of freedom is 
 \be
  \begin{split}
   {\rm spin-2}\,:&\qquad \frac{1}{2}n(n-1) -1\\
    {\rm spin-1}\,:&\qquad (n-1)(d-1) \\
    {\rm spin-0}\,:&\qquad \frac{1}{2}d(d+1)-d=\frac{1}{2}d(d-1)
  \end{split} 
 \ee
which sum up in total to 
 \be
  \frac{1}{2}(n+d)(n+d-3) \;, 
 \ee 
which equals  the number of degrees of freedom of a massless spin-2 field in $D=n+d$ dimensions. 
This spectrum agrees with  \cite{Han:1998sg,Giudice:1998ck}.

One  might  wonder about the second subsidiary condition in the first line of (\ref{FINALDYNEQ}), 
which mixes  $\underline{h}$ and $\underline{\varphi}$. There is nothing wrong with this,  but we may perform the following on-shell field redefinition to diagonalize 
the equations. There is an \textit{on-shell projector}  
to divergence-free tensors: 
 \be
  P_{\mu\nu} := \eta_{\mu\nu} + K\partial_{\mu}\partial_{\nu}\;. 
 \ee
On non-zero mode fields obeying $\square +\Delta=0$ we have 
 \be
  P_{\mu}{}^{\rho} P_{\rho\nu} = P_{\mu\nu} \;, \qquad P_{\mu}{}^{\nu} \partial_{\nu} = 0  
  \;, \qquad P_{\mu}{}^{\mu} = n-1\;. 
 \ee 
Then setting 
 \be
  \underline{h}'_{\mu\nu} = \underline{h}_{\mu\nu} + \frac{1}{n-1} P_{\mu\nu}\underline{\varphi} \;,  
 \ee 
while leaving ${a}_{\mu m}$ and $\varphi_{mn}$ unchanged, we have 
\be
 (\square +\Delta)\underline{h}'_{\mu\nu} = 0\;, \qquad\; \,  \partial^\mu \underline{h}'_{\mu\nu} = 0\;, \qquad\;\; \underline{h}' = 0\;, 
\ee
with the other two equations in (\ref{FINALDYNEQ}) staying the same. In this form the massive spin-2 
equations take the standard Fierz-Pauli form.

\section{Non-linear Kaluza-Klein theory via homotopy transfer}

In this section we review and develop the techniques from homotopy algebras needed  to 
define gauge invariant fields to any order in perturbation 
theory.\footnote{We thank Christoph Chiaffrino for helpful conversations on the topics of this section.} 
In the first subsection we give an in principle self-contained  review of $L_{\infty}$ algebras and homotopy transfer, 
although for all proofs we refer to the literature \cite{Arvanitakis:2020rrk,Chiaffrino:2020akd,Chiaffrino:2023wxk}.  In the second subsection we define the gauge invariant field  variables, 
and in the third subsection we apply this to Kaluza-Klein theory on a torus.

\subsection{$L_{\infty}$ algebras and homotopy transfer }

An $L_{\infty}$ algebra is a structure on an integer graded vector space $X=\oplus_{i\in\mathbb{Z}}X_i$ such as (\ref{KKComplex}) 
given by a differential $\partial\equiv b_1:X_i\rightarrow X_{i+1}$ satisfying $\partial^2=0$, together 
with higher multilinear maps or brackets $b_n:X^{\otimes n}\rightarrow X$ for $n\geq 2$, which have $n$ inputs and one output 
in the graded vector space $X$ and are subject to certain relations. 
Specifically, the $b_n$ are of intrinsic degree $+1$, meaning the degree of its output is the sum of the 
degrees of the inputs plus one. Moreover, the $b_n$ are graded symmetric, so that, for instance, 
 \be
  b_n(x_1,x_2, \ldots, x_n) = (-1)^{x_1 x_2} b_n(x_2,x_1, \ldots, x_n) \ \in \ X_{\sum_{i=1}^{n}x_i+1}\;. 
 \ee
(We sometimes just write $x_i$ for its degree $|x_i|$ when it is clear from the context that this is what is meant.)  
More generally, changing the order of any two adjacent arguments of $b_n$ gives a sign whenever both arguments 
are of odd degree. 

Most importantly, the $b_n$ are subject to an infinite tower of generalized Jacobi identities, of which we display 
the first three: 
 \be\label{GENJacobi} 
  \begin{split}
   0 \ &= \ b_1^2(x)\;, \\
   0 \ &= \  b_1(b_2(x_1, x_2)) + b_2(b_1(x_1), x_2) + (-1)^{x_1}b_2(x_1,b_1(x_2))\;, \\
   0 \ &= \ b_2(b_2(x_1,x_2),x_3) + (-1)^{x_2x_3}b_2(b_2(x_1,x_3),x_2) + (-1)^{x_1(x_2+x_3)} b_2(b_2(x_2,x_3), x_1) \\
   \ &\quad +b_1(b_3(x_1,x_2,x_3)) \\
     \ &\quad + b_3(b_1(x_1), x_2, x_3) + (-1)^{x_1}b_3(x_1, b_1(x_2),x_3) +(-1)^{x_1+x_2}b_3(x_1,x_2,b_1(x_3))\;. 
  \end{split} 
 \ee
The first relation  makes $X$ into a chain complex with differential $\partial=b_1$. The second relation states that the 
differential acts via the Leibniz rule on the 2-bracket $b_2$ (with unconventional signs that are due to the so-called $b$-picture 
conventions that, however,  will be more convenient  in the following). Finally, the third relation encodes the failure of  the graded Jacobi identity 
for $b_2$ in terms of the differential and the 3-bracket $b_3$.

Any  (semi-classical) field theory can be encoded in an $L_{\infty}$ algebra, as displayed in the previous section 
 for the quadratic part 
of gravity on flat space. Interactions, higher order corrections to the gauge transformations and Noether identities, etc., 
are encoded in the higher brackets $b_n$. For instance, the non-linear equations of motion take the form of generalized
Maurer-Cartan equations for fields $\Phi$ living in degree zero: 
 \be\label{EOMMaurerCARTAN}
  b_1(\Phi)  +\frac{1}{2}b_2(\Phi,\Phi) + \frac{1}{3!} b_3(\Phi,\Phi,\Phi) + \cdots =0\;, 
 \ee
 while the non-linear gauge transformations can be written as 
  \be\label{nonlingauge23}
   \delta_{\Lambda}\Phi = b_1(\Lambda) + b_2(\Lambda, \Phi) + \frac{1}{2}b_3(\Lambda,\Phi,\Phi) + \frac{1}{3!} b_4(\Lambda,\Phi,\Phi,\Phi) + \cdots \;, 
  \ee
where $\Lambda\in X_{-1}$. Note that the outputs of each term in (\ref{EOMMaurerCARTAN}) live in degree one, in agreement 
with this space being referred to as the `space of equations of motion', while the output of each term in (\ref{nonlingauge23}) lives 
in degree zero, the space of fields, as it should be since this is an infinitesimal field transformation. 
Consistency conditions of field theory, such as covariance of the field equations (\ref{EOMMaurerCARTAN}) under gauge 
transformations (\ref{nonlingauge23}), then follow as a consequence of the generalized Jacobi identities (\ref{GENJacobi}). 

It must be emphasized that the $b_n$ in (\ref{EOMMaurerCARTAN}) with all arguments in degree zero are in principle unrelated 
to the $b_n$ in (\ref{nonlingauge23}), where one argument is in degree $-1$. Indeed, for gravity on flat space, at least  in its standard formulation,  
all $b_n$ in (\ref{EOMMaurerCARTAN}) are non-zero, corresponding to the non-polynomial interactions of the Einstein-Hilbert action, 
while the $b_n$ in (\ref{nonlingauge23}) for $n\geq 3$ are zero, corresponding to the standard action of the diffeomorphism Lie algebra.

For our applications below it will be important to have a more efficient formulation of $L_{\infty}$ algebras, 
in order to be able to prove statements to all orders in perturbation theory. In particular, we need a closed form characterization 
of the $L_{\infty}$ generalized Jacobi identities. Such a formulation exists in terms of the operator 
 \be\label{DSum} 
  D := b_1+b_2+b_3+b_4+ \cdots 
 \ee
that obeys $D^2=0$ if and only if the $b_n$ define an $L_{\infty}$ algebra. Of course, as written, $D$ does not make sense as its 
first term takes one input, its second term takes two inputs, etc.~However, one can make sense of it by working on a larger 
space, the symmetric algebra.

\subsubsection*{Coalgebra and coderivations }

The symmetric algebra $S(X)$ of a vector space $X$ consists of words of vectors in $X$, i.e., arbitrary strings 
$x_1\wedge x_2\wedge\cdots \wedge x_n$ with `letters'  $x_i\in X$. For ease of notation we will typically leave out the wedge symbol, 
with the understanding that all words are graded symmetric, e.g., 
 \be
  x_1x_2 = (-1)^{x_1x_2}x_2x_1  \;.  
 \ee
More formally, we can write the symmetric algebra $S(X)$ as the direct sum 
\be
 S(X) = \mathbb{R}\oplus X\oplus S^2(X)\oplus S^3(X)\oplus\cdots \;, 
\ee
where it is convenient to include the numbers $\mathbb{R}$. Thus, elements of $S(X)$ are words with zero 
letters (numbers), one letter (vectors in $X$), two letters (vectors in $S^2(X)$), etc. 
As the actual $L_{\infty}$ structure lives on the subspace $X$ it is useful to define an explicit projection map  
\be\label{pi1MAP} 
 \pi_1: \,S(X)\,\rightarrow\, X\;, 
\ee
that projects any (linear combination of) words onto its  linear part.\footnote{For instance, for $1+x+x_2x_3x_7\in \mathbb{R}\oplus X\oplus S^3(X)$ we 
simply have 
\be
 \pi_1(1+x+x_2x_3x_7) = x\;. 
\ee}

Importantly, the symmetric algebra $S(X)$ carries a product $\mu:S(X)\otimes S(X)\rightarrow S(X)$ 
and hence an algebra structure but also a coproduct $\Delta:S(X)\rightarrow S(X)\otimes S(X)$ and hence a coalgebra 
structure. The product simply puts the words together, e.g.,  
 \be
  \mu(x_1x_2\otimes x_3x_4x_5) 
   = x_1x_2x_3x_4x_5\;, \qquad
  \mu(1\otimes  x_1x_3)  =x_1x_3\;. 
 \ee
The coproduct is given by 
 \be\label{coproductDelta} 
 \begin{split} 
   \Delta(1) &= 1\otimes 1\;, \\
   \Delta(x) &= 1\otimes x+x\otimes 1\;, \\
   \Delta(x_1x_2) &= 1\otimes x_1x_2 +x_1x_2\otimes 1 
   +x_1\otimes x_2 + (-1)^{x_1x_2}x_2\otimes x_1 \;, \quad {\rm etc.} \;, 
 \end{split} 
 \ee
i.e.~is given by the graded symmetric sum over all possibilities  of decomposing a word into two factors.  
This coproduct is coassociative, which means it has the dual properties to a product being associative, 
but we will not display the details. 

In the following the coalgebra structure on $S(X)$ will be more fundamental, and we will need the notion of 
a \textit{coalgebra morphism} between two coalgebras. In our case this will be a map between the 
symmetric algebras of the vector spaces $X$ and $\mathring{X}$ related by homotopy transfer. 
A coalgebra morphism is then a map  $F:S(X) \rightarrow S(\mathring{X})$ that is compatible with the coalgebra structure. 
Concretely, $F$ is encoded in a series of degree zero graded symmetric multilinear maps $(f_1,f_2,f_3,\cdots)$, 
where $f_n:X^{\otimes n}\rightarrow \mathring{X}$, 
that commute with the coproduct in that 
 \be\label{coalgebramorph}
 \mathring{\Delta}F= (F\otimes F)\Delta\;, 
 \ee
where $\mathring{\Delta}$ is the coproduct on $S(\mathring{X})$, defined as for $S(X)$. 
Using (\ref{coproductDelta}), this condition fixes the action of the coalgebra morphism $F$. 
For instance, $F(1)=1$, and  for $1+x_1x_2\in \mathbb{R}\oplus S^2(X)$ we have 
  \be\label{morphismACtion} 
   F(1+x_1x_2) = 1+ f_2(x_1,x_2) + f_1(x_1)f_1(x_2) \ \in \ \mathbb{R}\oplus \mathring{X}\oplus S^2(\mathring{X}) \;, 
  \ee
which is quickly verified to be compatible with (\ref{coalgebramorph}). More general formulas follow similarly.   

We next display some important properties of the coproduct and of coalgebra morphisms. 
To this end we define the exponential map $\exp: X  \rightarrow S(X)$ by 
 \be\label{expMAP} 
  \exp(x) = \sum_{n\geq0} \frac{1}{n!} x^n = 1+ x + \frac{1}{2} x^2+\frac{1}{3!} x^3+\cdots\;, 
 \ee
where the products on the right-hand side are the (graded symmetric) wedge products  of the symmetric algebra $S(X)$. 
Note that acting with the projection (\ref{pi1MAP}) to the linear piece gives the identity,  
 $\pi_1\circ \exp={\rm id}_{X}$. Further note that for any element $x$ of odd degree we have $xx=-xx$ and hence $x^2=0$, so for odd elements 
$\exp(x)=1+x$. From the definition such  familiar relations as $\exp(x+y)=\exp(x)\cdot \exp(y)$ follow, 
but only \textit{provided} not both $x$ and $y$ are odd, because the product on the right-hand side 
is the wedge product of $S(X)$. Moreover, the coproduct acts on $\exp(x)$ for $x$ even   as 
 \be\label{DeltaExpProp} 
  \Delta \exp(x) = \exp(x) \otimes \exp(x) \;,  
 \ee
as one may verify by writing out both sides using (\ref{coproductDelta}). 
Finally, the action of  any coalgebra morphism $F=(f_1,f_2,f_3,\ldots)$, as displayed in (\ref{morphismACtion}),  is such that 
 \be\label{morhismEXP}
  F\exp(x) = \exp\big(\widetilde{f}(x)\big)\;, 
 \ee
 where we defined the non-linear map $\widetilde{f}=\pi_1\circ F\circ \exp :X\rightarrow X$ associated to $F$: 
  \be\label{morhismEXP2}
   \widetilde{f}(x) := 
   f_1(x) + \frac{1}{2} f_2(x,x) +\frac{1}{3!} 
   f_3(x,x,x) + \cdots \;. 
  \ee
Note that the non-linear terms are zero when $x$ is a homogenous element of  odd degree.   
Let us emphasize, however, that the above formulas  also make sense  when $x$ is not homogenous and so does not have a definite degree, say, 
when it is the sum of two vectors of different degrees, as will be needed below. 
It is again straightforward to verify (\ref{morhismEXP})  order by order.

We are now ready to define \textit{coderivations}. Given an algebra one defines a derivation as an operator 
that obeys the Leibniz rule with respect to the product. Similarly, given a coalgebra one defines a coderivation 
as an operator that obeys the naturally dual co-Leibniz rule  with respect to the coproduct.  
Without displaying the details we give the action for the sum of $L_{\infty}$ maps $D=\partial+b_2+b_3+\cdots $, c.f.~(\ref{DSum}), 
in terms of their projection $\pi_1D$ via (\ref{pi1MAP}) to the linear part, where $\pi_1D$ of course acts on any word consisting of $n$ letters 
by inserting the letters into $b_n$. 
The co-Leibniz rule extends this action to $S(X)$ by use of the coproduct and product: 
 \be\label{CoderLift} 
   D = \mu\circ  \big(\pi_1D\otimes 1\big)\circ \Delta \;. 
 \ee 
For instance, on $x\in S^1(X)=X$ one obtains  
 \be
  D(x) = \mu\big(\pi_1D\otimes 1\big)\Delta(x) =  \mu\big(\pi_1D\otimes 1\big)(1\otimes x+x\otimes 1) 
  =\mu\big(\partial (x)\otimes 1\big) = \partial(x)\;, 
 \ee 
where we used $D(1)=0$. 
Thus, as to be expected, on a single vector only the first term in $D=\partial+\cdots $ acts non-trivially. 
Next, on $x_1x_2\in S^2(X)$ one obtains 
 \be
  \begin{split} 
   D(x_1x_2) &=  \mu \big(\pi_1D\otimes 1\big)\big(1\otimes x_1x_2 +x_1x_2\otimes 1 
   +x_1\otimes x_2 + (-1)^{x_1x_2}x_2\otimes x_1\big)\\
   &= \mu\big(b_2(x_1,x_2)\otimes 1 +\partial(x_1)\otimes x_2+ (-1)^{x_1x_2} \partial(x_2) \otimes x_1\big) \\
   &= b_2(x_1,x_2) +\partial(x_1) x_2+ (-1)^{x_1x_2} \partial(x_2)  x_1 \\
   &= b_2(x_1,x_2) +\partial(x_1) x_2+ (-1)^{x_1} x_1\partial(x_2) \ \in \ X\oplus S^2(X) \;, 
  \end{split}
 \ee
using the graded symmetry in the last step.  
This  implies, upon considering the special case that $D=\partial$ just consists of a linear map, that  
 \be
  \partial( x_1x_2) = \partial(x_1)x_2+(-1)^{x_1} x_1\partial(x_2)\;, 
 \ee
which is the familiar Leibniz rule.  
Thus, for a linear map, being a coderivation  is the same as being a derivation. 
The symmetric algebra $(S(X), \partial)$ equipped with the differential $\partial$ acting as a 
derivation (or coderivation) is  in particular a chain complex. 
Finally, let us display the coderivation acting on a 3-letter word for the special case of $D:=b_2$ consisting  only of the bilinear map. 
By an analysis similar to the above one finds 
 \be\label{b2action} 
  b_2(x_1x_2x_3) = b_2(x_1,x_2)x_3+(-1)^{x_1(x_2+x_3)} b_2(x_2,x_3)x_1
  +(-1)^{x_2x_3 } b_2(x_1,x_3)x_2\;. 
 \ee
More generally, a coderivation $b_n: S^{m}(X)\rightarrow S^{m-n+1}(X)$ acts non-trivially only on words with $m\geq n$ letters 
by summing over all possibilities of picking out $n$ letters, keeping their order unchanged while  moving them to the 
front (including the sign factors obtained by doing so),  and then inserting the $n$ vectors into $b_n$.  
With these rules it is easy to verify that $D^2=0$ for $D=b_1+b_2+b_3+\cdots$, evaluated on linear, bilinear and trilinear objects, respectively, 
gives rise to the first three $L_{\infty}$ relations (\ref{GENJacobi}). 
 
We close this part by displaying a curious relation for the exponential map (\ref{expMAP}) acting on an even $x$, that will be needed 
below  and  that follows from  (\ref{CoderLift}):
 \be\label{strangeDProp}
  \begin{split}
    D \exp(x)  &= \mu(\pi_1D\otimes 1)\Delta \exp(x)   \\
   &=  \mu(\pi_1D\otimes 1)
   (\exp(x)\otimes \exp(x))\\
   &= \mu\big(\pi_1D\exp(x)
   \otimes
   \exp(x)\big) \\
   &= \pi_1D\exp(x)\cdot 
   \exp(x)\;, 
  \end{split}
 \ee
where we used (\ref{DeltaExpProp}) in the second line. Note that since (\ref{DeltaExpProp}) is only valid for homogeneous $x$ 
of even degree, the same is true for  (\ref{strangeDProp}).

\subsubsection*{Perturbation lemma and homotopy transfer} 
 
 We will now show that given homotopy data relating two  chain complexes $p: (X,\partial)\rightarrow (\mathring{X},\mathring{\partial})$, 
 an $L_{\infty}$ algebra on $X$ is transported to an $L_{\infty}$ algebra on $\mathring{X}$. 
 This is established by defining a coderivation $\mathring{D}$ on $S(\mathring{X})$, which squares to zero and hence defines 
 an $L_{\infty}$ algebra on $\mathring{X}$, in terms of the nil-potent coderivation $D$ that defines the $L_{\infty}$ algebra on $X$.
 
 To this end we need to uplift the homotopy data to the full symmetric algebras. For projection and inclusion these 
 act as morphisms in that 
 \be
 \begin{split} 
  \iota(\mathring{x}_1\cdots \mathring{x}_n) &= \iota(\mathring{x}_1) \cdots \iota(\mathring{x}_n)\;, \\
  p(x_1\cdots x_n) &= p(x_1)\cdots p(x_n)\;,  
 \end{split} 
 \ee
but for the homotopy $\frak{h}$ it is more subtle, requiring the so-called $(\iota p,{\bf 1})$ Leibniz rule. For instance, for  a word with two letters: 
 \be\label{Haction} 
  \frak{h}(x_1x_2) = \frac{1}{2}\big( \frak{h}(x_1)x_2 + (-1)^{x_1} \iota p(x_1)\frak{h}(x_2) +(-1)^{x_1x_2}\frak{h}(x_2)x_1 + (-1)^{(x_1+1)x_2} \iota p(x_2)\frak{h}(x_1)\big) \;.  
 \ee
More generally, $\frak{h}$ acts with the (graded symmetrized) Leibniz rule, except that $\iota p$ acts on all factors that have been jumped. 
With these lifts for $p,\iota, \frak{h}$, which for ease of notation we denote by the same letters, 
the homotopy relations (\ref{homotopyREL}) hold on the entire symmetric algebra.

In order to define the coderivation $\mathring{D}$ on $S(\mathring{X})$ we will employ the so-called perturbation lemma. 
The data and assumptions required by the perturbation lemma are satisfied  here: we have two chain complexes $(S(X), \partial)$ and $(S(\mathring{X}),\mathring{\partial})$ with homotopy 
data $p: (S(X), \partial)\rightarrow (S(\mathring{X}),\mathring{\partial})$, $\iota: (S(\mathring{X}),\mathring{\partial})\rightarrow (S(X), \partial)$ and 
$\frak{h}: (S(X), \partial)\rightarrow (S(X), \partial)$ obeying the right  relations. 
We then view the non-linear $L_{\infty}$ brackets as a perturbation of the differential $\partial=b_1$ that encodes  the free theory:  
 \be
  D:= \partial + B\;, \qquad B=b_2+b_3+b_4+\cdots\;, 
 \ee
where $B$ is a coderivation, so that $D^2=0$. 
The perturbation lemma now states that 
 \be\label{transportedD} 
  \mathring{D} : = \mathring{\partial} +p(B-B\frak{h}B+ B\frak{h}B\frak{h}B+\cdots)\iota 
 \ee
still satisfies $\mathring{D}^2=0$ and hence defines an $L_{\infty}$ algebra on $\mathring{X}$. 
Furthermore, there are deformations for all homotopy data: 
 \be\label{HomotopyDEF}
  \begin{split}
   P&= p\big(1-B\frak{h} + B\frak{h}B\frak{h} - \cdots\big)\;,  \\
   I&=\big(1-\frak{h}B+\frak{h}B\frak{h}B -  \cdots\big)\iota\;,  \\
    \frak{H}&= \frak{h}\big(1-B\frak{h}+B\frak{h}B\frak{h} -  \cdots\big) \;, 
  \end{split}
 \ee
so that the homotopy relations hold: 
 \be\label{liftedhomotopy} 
  IP+D\frak{H}+\frak{H}D={\rm id}_{S(X)} \;, \qquad 
  PI={\rm id}_{S(\mathring{X})} \;. 
 \ee
Assuming that the side conditions (\ref{sideconditions}) are satisfied, 
$P$ is a  coalgebra morphism and a chain map: 
 \be\label{chainMapCOND}
  \mathring{D} P = P{D} \;. 
 \ee

\medskip

Let us spell out explicitly  the homotopy transported 2- and 3-brackets.   For the 2-bracket one computes 
with (\ref{transportedD}) by acting on a quadratic monomial: 
 \be
  \mathring{b}_2(\mathring{x}_1, \mathring{x}_2) = \pi_1\mathring{D}(\mathring{x}_1\mathring{x}_2) 
  =\pi_1pB\iota (\mathring{x}_1\mathring{x}_2) = p\big(b_2(\iota(\mathring{x}_1),\iota(\mathring{x}_2))\big)\;, 
 \ee
where we used that, thanks to the linear projection $\pi_1$, only the first term in $B=b_2+\cdots$ 
acts non-trivially. 
In the following we will sometimes use  the notation $x_1=\iota(\mathring{x}_1)$, etc., when there is no danger of confusion, 
so that the above reads 
 \be\label{transported2Bracket} 
   \mathring{b}_2(\mathring{x}_1, \mathring{x}_2)= p\big(b_2({x}_1,{x}_2)\big)\;. 
 \ee
In words, the transported 2-brackets is given by evaluating the original $b_2$ on the included arguments  
and then projecting down to $\mathring{X}$. 

Next, we compute the transported 3-bracket 
using the same notation. In (\ref{transportedD}) we now have to include the next term in the geometric series: 
 \be
  \begin{split}
\mathring{b}_3(\mathring{x}_1,\mathring{x}_2,\mathring{x}_3) &= \pi_1\mathring{D}  (\mathring{x}_1\mathring{x}_2\mathring{x}_3)\\
&=\pi_1p\big(B(x_1x_2x_3) -B\frak{h}B(x_1x_2x_3)\big)\\
&=p(b_3(x_1,x_2,x_3))\\
&\; - p\big(B\frak{h}\big(b_2(x_1x_2)x_3
+(-1)^{x_1(x_2+x_3)} b_2(x_2x_3)x_1+(-1)^{x_2x_3} b_2(x_1x_3)x_2\big)\big)\,, 
  \end{split}
 \ee
where we used (\ref{b2action}).  
To evaluate the action of $\frak{h}$ in the last  line we recall that the side conditions (\ref{sideconditions})  imply with our notation 
$\frak{h}(x_1)=\frak{h}\iota(\mathring{x_1})=0$ 
and $\iota p(x_3)=\iota p\iota (\mathring{x}_3)=\iota (\mathring{x}_3)=x_3$
so that the homotopy action (\ref{Haction})  reduces to 
 \be
  \frak{h}(b_2(x_1,x_2)x_3) = \frak{h}(b_2(x_1,x_2))x_3\;, 
 \ee
and similarly for the other terms. Therefore, 
 \be\label{transport3Bracket} 
   \begin{split}
&\mathring{b}_3(\mathring{x}_1,\mathring{x}_2,\mathring{x}_3) =
 p\Big\{ b_3(x_1,x_2,x_3)\\
&- b_2(\frak{h}(b_2(x_1,x_2)),x_3) 
-(-1)^{x_1(x_2+x_3)}b_2(\frak{h}(b_2(x_2,x_3)),x_1) -(-1)^{x_2x_3} b_2(\frak{h}(b_2(x_1,x_3)),x_2)\Big\} \,. 
  \end{split}
 \ee
Higher transported brackets are similarly obtained by nesting of the $b_n$ with suitable insertions of the 
homotopy map $\frak{h}$ (which makes the degrees consistent, with $b_n$ having degree $+1$ and $\frak{h}$ 
having degree $-1$). Such expressions for the transported brackets can be given a graphical representation 
in terms of tree-diagrams that is entirely analogous to the Feynman diagrams of tree-level scattering amplitudes, 
with $\frak{h}$ playing the role of the Feynman propagator. In fact, the computation of tree-level 
scattering amplitudes can also be interpreted as homotopy transfer, see \cite{Bonezzi:2023xhn} for a self-contained 
review.

\subsection{Gauge invariant  fields} 

We will now apply the homotopy algebra techniques reviewed in the previous subsection to define 
gauge invariant field variables to all orders in perturbation theory. We begin by writing the 
full non-linear gauge transformations (\ref{nonlingauge23}) in terms of the coderivation $D$, using the 
exponential map (\ref{expMAP}): 
 \be\label{originalGauge} 
  \delta_{\Lambda}\Phi =  \pi_1D\Big[\exp(\Phi+\Lambda)\Big]' \;, 
 \ee
where the notation $\big[\;\;\big]'$ indicates that in the exponential  series one picks out 
precisely the terms that are linear in $\Lambda$.\footnote{Rescaling $\Lambda\rightarrow \varepsilon\Lambda$ 
one could write this as $\big[\;\;\big]'=\frac{d}{d\varepsilon}\big[\;\,\big] \big|_{\varepsilon=0}$,  but  it should be clear enough what 
it means to pick out the terms linear in $\Lambda$. Note that since $\Lambda$ is of odd degree, $\Lambda^2=0$ and 
so only the zeroth order terms  are truncated by $\big[\;\;\big]'$.}
Then, indeed, 
 \be
 \begin{split}
   \delta_{\Lambda}\Phi &= \pi_1D \Big[\Phi+\Lambda +\frac{1}{2} (\Phi+\Lambda)^2
   +\frac{1}{3!} (\Phi+\Lambda)^3+\cdots \Big]' \\
   &=  \pi_1\big(b_1+b_2+b_3+\cdots\big)  \Big(\Lambda  + \Phi\Lambda +\frac{1}{2}  \Phi\Phi\Lambda +\cdots  \Big)\\
   &= b_1(\Lambda) + b_2(\Phi,\Lambda) + \frac{1}{2} b_3(\Phi,\Phi, \Lambda)  +\cdots\;, 
\end{split} 
 \ee
in agreement with (\ref{nonlingauge23}).

It is useful to also  note that \textit{on-shell} 
one may write the gauge transformations more simply, without the projections $\pi_1$ and $\big[\;\;\big]'$, as  
 \be\label{onshellgauge} 
  \delta_{\Lambda}\exp(\Phi) 
  \doteq D \Big[\exp(\Phi+\Lambda)
  \Big]\;, 
 \ee
where $\doteq$ indicates that this relation only holds on-shell, i.e., provided the Maurer-Cartan equations 
(\ref{EOMMaurerCARTAN}) are satisfied. Indeed, expanding the right-hand side of (\ref{onshellgauge})
 one also picks up terms like $b_1(\Phi)\Lambda+\cdots$, which do not appear in the 
gauge variations on the left-hand side but are zero on-shell. 
More precisely, we first note that the Maurer-Cartan equations (\ref{EOMMaurerCARTAN}) can be written in terms of the coderivation $D$ as 
\be
 \pi_1D\exp(\Phi) = 0\;, 
\ee 
as follows quickly by expansion.   By (\ref{strangeDProp}) the Maurer-Cartan equations are 
then equivalent to $D\exp(\Phi)=0$: 
 \be\label{MCEquivalence} 
  \pi_1D\exp(\Phi)=0  \quad 
  \Leftrightarrow \quad D\exp(\Phi)=0\;. 
 \ee 
We thus have 
 \be
 D \Big[\exp(\Phi+\Lambda)\Big] = D\Big[\exp(\Phi)\cdot \exp(\Lambda)\Big] = D\Big[\exp(\Phi) + \exp(\Phi)  \Lambda\Big] 
 \doteq D\Big[\exp(\Phi)  \Lambda\Big]\;, 
 \ee
where we used $\exp(\Lambda)=1+\Lambda$ following from $\Lambda$ having odd degree. Moreover, we assumed the Maurer-Cartan equation in the last step. 
Expanding out the right-hand side in here one may confirm that this equals  $\delta_{\Lambda}\exp(\Phi)$ for  (\ref{nonlingauge23}), 
thereby proving (\ref{onshellgauge}).

We now claim that the gauge invariant (or, more precisely, gauge covariant) field variable is given by 
the action of the non-linearly corrected projector in (\ref{HomotopyDEF}) on the exponential: 
  \be\label{gaugecovField}
 \begin{split} 
  \widehat{{\Phi}} &:= \pi_1P\exp(\Phi)\;. 
 \end{split}
 \ee
Recalling that $P$ is a coalgebra morphism, we can use (\ref{morhismEXP}) to write 
$P\exp(\Phi) = \exp(\widetilde{p}(\Phi))$, so that we can also write  
 \be\label{PhiHatal} 
  \widehat{{\Phi}} = \widetilde{p}(\Phi) = p_1(\Phi)+\frac{1}{2}p_2(\Phi,\Phi) + \frac{1}{3!}p_3(\Phi,\Phi,\Phi)+\cdots\;, 
 \ee
where $(p_1,p_2,p_3,\ldots)$ are the multilinear maps defining the morphism $P$. 
Moreover, $\widehat{\Phi}$ can be defined indirectly without projection $\pi_1$ by writing  
 \be\label{expphihat} 
  \exp(\widehat{\Phi})= P \exp(\Phi) \;. 
 \ee

Next, we show that $\widehat{\Phi}$ is gauge covariant. This is easiest shown using (\ref{expphihat}) and 
 the on-shell form (\ref{onshellgauge})  of the gauge transformations: 
  \be\label{covGAUGE233}
  \begin{split} 
   \delta_{\Lambda} \exp(\widehat{\Phi})&= P  \,\delta_{\Lambda} \exp(\Phi) 
   \doteq P D \Big[\exp(\Phi+\Lambda) \Big] \\
   &= \mathring{D} P  \Big[\exp(\Phi+\Lambda)
  \Big] 
   = \mathring{D}   \Big[\exp\big(\widetilde{p}(\Phi+\Lambda)\big)
  \Big] \\
  &= \mathring{D}\Big[\exp(\widehat{\Phi}+\widehat{\Lambda}(\Lambda,\Phi)) \Big] \;. 
  \end{split} 
  \ee
Here we used in the second  line (\ref{chainMapCOND})  and (\ref{morhismEXP}). 
In the last line we used 
  \be\label{widehatLambda}
  \begin{split} 
\widetilde{p}(\Phi+ \Lambda) &= 
  p_1(\Phi+\Lambda) 
  +\frac{1}{2}p_2
  (\Phi+\Lambda,\Phi+\Lambda) +\frac{1}{3!} p_3(\Phi+\Lambda,\Phi+\Lambda,\Phi+\Lambda) +\cdots \\
  &=  p_1(\Phi) + \frac{1}{2}p_2(\Phi,\Phi) + \frac{1}{3!} p_3(\Phi,\Phi,\Phi)+\cdots \\
   &\quad  + 
     p_1(\Lambda)+p_2(\Phi, \Lambda)+\frac{1}{2}p_3(\Phi,\Phi,\Lambda) +  \cdots \\
  &= 
 \widehat{\Phi}  +
  \widehat{\Lambda}(\Lambda,\Phi)\;, 
  \end{split} 
  \ee
where we recalled (\ref{PhiHatal}) and used that all contributions of higher order in $\Lambda$ vanish due 
to $\Lambda$ having odd degree. This defines a new (field dependent) gauge parameter: 
 \be\label{LambdaHAT} 
   \widehat{\Lambda}(\Lambda,\Phi) :=  p_1(\Lambda)+p_2(\Phi, \Lambda)+\frac{1}{2}p_3(\Phi,\Phi,\Lambda)  + \cdots\;. 
 \ee
The gauge transformation (\ref{covGAUGE233})  takes the same form as (\ref{onshellgauge}), except with 
$D$ being replaced by $\mathring{D}$, $\Phi$ being replaced by $\widehat{\Phi}$ and the gauge parameter 
being replaced by $ \widehat{\Lambda}(\Lambda,\Phi)$. Thus, with respect to this new gauge parameter (\ref{LambdaHAT}), 
the field $\widehat{\Phi}$ transforms covariantly according to the $L_{\infty}$ algebra on $\mathring{X}$. 
In particular, since $\mathring{X}$ carries only zero mode gauge parameters, $\widehat{\Phi}$ is fully gauge invariant 
under all non-zero mode gauge transformations. 

The above proof established gauge invariance of $\widehat{\Phi}$, but only on-shell. We can, however, argue as follows that 
gauge invariance holds also off-shell, assuming only  that the gauge algebra closes off-shell (as is the case in Einstein gravity 
and exceptional field theory). To this end we note that under the assumption of off-shell closure  one obtains an 
$L_{\infty}$ algebra on the sub-complex that truncates out field equations and higher spaces (for instance, the pure gravity 
complex (\ref{KKComplex}) would be truncated to $X_{-1}\rightarrow X_0$). This algebra, called $L_{\infty}^{\rm gauge+fields}$ in \cite{Hohm:2017pnh}, 
encodes the non-linear gauge transformation and  their closure. On this subcomplex, the Maurer-Cartan equations are trivially 
satisfied, for degree reasons, and hence the above proof goes through. 

It would be reassuring, however, to verify gauge invariance on the full complex, which is possible upon a further assumption.  
We compute the gauge variation of (\ref{gaugecovField}) using (\ref{originalGauge}): 
 \be\label{hattedgaugecovariance} 
 \begin{split}
  \delta_{\Lambda}\widehat{\Phi}
  &=\pi_1P \,\delta_{\Lambda}
  \exp(\Phi) 
  =\pi_1P\, \delta_{\Lambda}\Phi 
  \exp(\Phi) \\
  &=\pi_1P\, \pi_1D\Big[\exp(\Phi+\Lambda)\Big]' \exp(\Phi) 
  \\
  &= \Big[\pi_1P\,\pi_1 D \exp(\Phi+\Lambda)\cdot \exp(\Phi) \Big]'
  \\
   &= \Big[\pi_1P\, \pi_1 D \exp(\Phi+\Lambda)\cdot \exp(\Phi+\Lambda) \Big]'
  \\
  &= \Big[  \pi_1P D \exp(\Phi+\Lambda)\Big]'
  \\
  &= \Big[  \pi_1 \mathring{D}P  \exp(\Phi+\Lambda)\Big]'
  \\
  &= \Big[  \pi_1 \mathring{D}  \exp\big(\widehat{\Phi} + \widehat{\Lambda}(\Lambda,\Phi)\big)  \Big]'\;. 
 \end{split}
 \ee
Here the step in the fourth  line holds under an assumption on $P$ to be discussed below; 
in the fifth  line we used (\ref{strangeDProp});\footnote{Strictly speaking, (\ref{strangeDProp}) is not immediately applicable as it was derived only for homogenous arguments  of even degree, but one may convince oneself that the failure acting on $\Phi+\Lambda$ is of higher order 
in $\Lambda$ and hence drops out under the outer 
projector $\big[\;\;\big]'$.}
 in the sixth  line we used (\ref{chainMapCOND}); the last step follows as in (\ref{covGAUGE233}), with the same effective 
 parameter (\ref{LambdaHAT}). The final gauge variation takes the same form as (\ref{originalGauge}), except with 
 respect to the $L_{\infty}$ algebra on $\mathring{X}$. This means  that $\widehat{\Phi}$ transforms covariantly 
 as in (\ref{nonlingauge23})  under the gauge transformations governed by the transported $L_{\infty}$ algebra on $\mathring{X}$, i.e., 
   \be\label{nonlingauge237}
   \delta_{\Lambda}\widehat{\Phi} = \mathring{b}_1\big(\widehat{\Lambda}\big) 
   + \mathring{b}_2\big(\widehat{\Lambda}, \widehat{\Phi}\big) + \frac{1}{2}\mathring{b}_3
   \big(\widehat{\Lambda},\widehat{\Phi},\widehat{\Phi}\big)
    + \frac{1}{3!} \mathring{b}_4\big(\widehat{\Lambda},\widehat{\Phi},\widehat{\Phi},\widehat{\Phi}\big) + \cdots \;. 
  \ee
 In particular, since $\widehat{\Lambda}$ in (\ref{LambdaHAT}) belongs to $\mathring{X}$, hence carrying  
 only zero modes, it follows that $\widehat{\Phi}$ 
 is strictly  gauge invariant under all non-zero mode gauge parameters. 
 
 It remains to justify the assumption on $P$ used above. In the step from the third to the fourth line in (\ref{hattedgaugecovariance}) 
 we added the terms $\big[\pi_1P\, \pi_1 D \exp(\Phi+\Lambda)\cdot \exp(\Phi)\cdot \Lambda \big]'$ that are generally non-zero, 
 but are actually zero for a large class of $L_{\infty}$ algebras including the Kaluza-Klein example discussed here. 
 Given the outer projection $\big[\;\;\big]'$ to terms linear in $\Lambda$ a potentially dangerous new term would arise when 
 the coderivation acts on $\Phi$, leading to lowest order to the new term $\pi_1P(b_1(\Phi)\Lambda)$.  Note that with $\Lambda$ having degree $-1$
 and $b_1(\Phi)$ having degree $+1$ this is indeed a term of degree zero, as required for the variation of a field, 
 and so this term can exist by degree reasons. However, for the projector (\ref{HomotopyDEF}) and homotopy map $\frak{h}$ considered here 
 this term is actually zero. This follows from $\frak{h}$ being trivial in degree $+1$ and $-1$: 
 \be\label{funnycontribu} 
  \pi_1P(b_1(\Phi)\Lambda) 
  =\pi_1p(1-B\frak{h}+\cdots)(b_1(\Phi)\Lambda)
  =-\pi_1pB\frak{h}(b_1(\Phi)\Lambda) = 0 \;. 
 \ee
Indeed, with the action (\ref{Haction}) of the homotopy map $\frak{h}$ this gives zero due to $\frak{h}_{1}=0=\frak{h}_{-1}$, 
see (\ref{homotopy0Torus3}).  
More generally, given these properties of $\frak{h}$, we claim that all non-linear projection maps 
evaluated on one degree $1$ and one degree $-1$ argument are  zero:
 \be
  p_m(b_n(\Phi,\ldots,\Phi),\Phi, \ldots,\Phi, \Lambda) =0\;. 
 \ee
 This implies that the terms added in the above derivation are in fact zero.

 For future applications let us explore potential contributions such as (\ref{funnycontribu}) in more detail. 
 Suppose we had gauge for gauge symmetries. Then, in general, we would also have a homotopy map $\frak{h}_{-1}$ and thus, 
 from the right-hand side of (\ref{funnycontribu}), contributions like $b_2(b_1(\Phi), \frak{h}_{-1}(\Lambda))$ pairing 
 equations of motion and gauge for gauge parameters. Such brackets exist if and only if the gauge for gauge property 
 holds only on-shell. Indeed, up to and including linear terms in fields,  the trivial gauge parameter is of the form 
  \be
   \Lambda_{\rm triv} := b_1(\chi)+ b_2(\chi,\Phi)+\cdots\;, \qquad \chi\in X_{-2}\;, 
  \ee
 for then the gauge transformation (\ref{nonlingauge23}) reads 
  \be
  \begin{split} 
   \delta_{ \Lambda_{\rm triv} }\Phi  &= b_1(b_1(\chi)+b_2(\chi,\Phi))  + b_2(b_1(\chi),\Phi)+\cdots \\
   &= -b_2(\chi, b_1(\Phi)) + \cdots \;, 
  \end{split} 
  \ee
where we used the $L_{\infty}$ relations (\ref{GENJacobi}). We thus infer that in general $ \chi\in X_{-2}$ only gives rise 
to a gauge transformation that is zero on-shell, i.e., provided $b_1(\Phi)+\cdots=0$. 
Assuming that the $L_{\infty}$ brackets do not pair a trivial parameter 
with field equations thus amounts to assuming that the gauge for gauge property holds off-shell, as is indeed 
the case in a large class of theories including exceptional field theory. For these, contributions like 
$b_2(b_1(\Phi), \frak{h}_{-1}(\Lambda))$  do not enter. 
In contrast, contributions like $b_2(\frak{h}_1(b_1(\Phi)), \Lambda)$ will enter once equations of motion are partially solved 
so that there will be a non-trivial $\frak{h}_1$. But this is as to be expected because solving equations of motion requires partial 
gauge fixing (in the Kaluza-Klein case of the zero-mode diffeomorphisms), which in turn changes 
the notion of covariance. Of course, this will not affect the fact that the $\widehat{\Phi}$ are fully gauge invariant under all 
non-zero mode gauge transformations.

 As a consistency check we will next verify explicitly, to second order in fields, that (\ref{gaugecovField}) indeed yields a gauge invariant field variable. 
 Expanding the projector $P$ and the exponential map we have 
 \be
 \begin{split}
  \widehat{\Phi} &= \pi_1p\big(1-B\frak{h}+\cdots\big)\big(1+\Phi+\tfrac{1}{2}\Phi^2+\cdots\big) \\
  &=  p(\Phi) -\frac{1}{2} pb_2(\frak{h}(\Phi),\Phi)
  -\frac{1}{2} pb_2(\frak{h}(\Phi), \iota p(\Phi))\\
  &\equiv  p_1(\Phi)  +\frac{1}{2} p_2(\Phi,\Phi)+\cdots \;, 
 \end{split}
 \ee
using $\pi_1(1)=0$. From this we infer $p_1=p$ and 
 \be
 \begin{split}
  p_2(\Phi,\Phi) &= -pb_2(\frak{h}(\Phi),\Phi)
  -pb_2(\frak{h}(\Phi), \iota p(\Phi)) \\
  &=-2 p b_2(\frak{h}(\Phi), \Phi)  +pb_2(\frak{h}(\Phi), 
  \partial(\frak{h}(\Phi)))\;, 
 \end{split} 
 \ee
where we used  the homotopy relation $\iota p (\Phi) = \Phi - \partial (\frak{h} \Phi)$ in the second line. 
This determined  the `polarized' $p_2$ on two degree zero arguments $\Phi$ that are in fact the same. 
Evaluated on two arbitrary degree zero arguments it takes the symmetrized  form 
  \be\label{p2unpolarized}
 \begin{split}
  p_2(\Phi_1,\Phi_2) 
  = &- p b_2(\frak{h}(\Phi_1), \Phi_2)- p b_2(\frak{h}(\Phi_2), \Phi_1) \\
   & +\frac{1}{2} pb_2(\frak{h}(\Phi_1), \partial(\frak{h}(\Phi_2)))+\frac{1}{2} pb_2(\frak{h}(\Phi_2), \partial(\frak{h}(\Phi_1)))\;. 
 \end{split} 
 \ee 

We next compute the gauge transformation of $\widehat{\Phi}$ up to and including linear order in $\Phi$: 
 \be\label{COVAriance1} 
  \begin{split}
    \delta_{\Lambda}\widehat{\Phi} &= 
    p(\delta_{\Lambda}\Phi)  
    + p_2(\delta_{\Lambda}\Phi,\Phi)
    +\cdots \\
    &=p\big(\partial({\Lambda})+
    b_2(\Lambda, \Phi)\big)+ p_2(\partial(\Lambda), \Phi) +\cdots \\
    &=\mathring{\partial}(p({\Lambda}))+
     p(b_2(\Lambda, \Phi))+p_2(\partial(\Lambda), \Phi) +\cdots 
  \end{split}
 \ee
where we used the chain map property for $p$ in the first term.  
Covariance means that this should be equal to 
\be\label{COVAriance2} 
 \begin{split}
    \delta_{\Lambda}\widehat{\Phi} & \overset{!}{=}
    \mathring{\partial}\big(\widehat{\Lambda}(\Lambda,\Phi)\big) + \mathring{b}_2\big(\widehat{\Lambda}(\Lambda,\Phi),
    \widehat{\Phi}\big) +\cdots  \\
    &= \mathring{\partial}(p(\Lambda)+p_2(\Lambda, \Phi)) 
    + pb_2\big(\iota p({\Lambda}), \iota p({\Phi})\big) + \cdots\;,  
   \end{split}
 \ee
with respect to the new gauge parameter $\widehat{\Lambda}(\Lambda,\Phi)=p(\Lambda)+p_2(\Phi,\Lambda)+\cdots$ 
in (\ref{LambdaHAT}) that we wrote 
out in the second line. The zeroth order terms in (\ref{COVAriance1}) and (\ref{COVAriance2}) agree. 
To linear order in $\Phi$, covariance amounts to 
 \be\label{COVariancecond}
   p\big\{b_2(\Lambda, \Phi)
  -b_2(\iota p({\Lambda}), \iota p({\Phi}))\big\} 
  \overset{!}{=} \mathring{\partial} p_2(\Lambda, \Phi)
  - p_2(\partial\Lambda, \Phi) \;. 
 \ee
With the homotopy relations $\iota p (\Lambda) = \Lambda - \frak{h}(\partial \Lambda)$ and $\iota p (\Phi) = \Phi - \partial (\frak{h} \Phi)$, 
where one uses $\frak{h}_{-1}(\Lambda)=0$ and $\frak{h}_1(\partial\Phi)=0$,
we can rewrite the terms on the left-hand side as  
 \be
 \begin{split}
  b_2(\Lambda, \Phi)
  -b_2(\iota p({\Lambda}), \iota p({\Phi}))
  =b_2(\Lambda, \partial(\frak{h}\Phi)) + 
  b_2(\frak{h}(\partial\Lambda), \Phi) 
  -b_2(\frak{h}(\partial\Lambda), \partial(\frak{h}\Phi)) \;. 
 \end{split} 
 \ee
We now claim that this equals the right-hand side of  (\ref{COVariancecond}) for 
 \be\label{p2pnlambdaphi} 
   p_2(\Lambda, \Phi) = 
 - p b_2(\frak{h}\Phi, \Lambda) - \frac{1}{2}pb_2(\frak{h}(\partial\Lambda), \frak{h}\Phi)\;. 
 \ee
One may verify that this expression agrees with the definition of  the perturbed projector 
(\ref{HomotopyDEF}) to linear order in fields. 
Acting now on (\ref{p2pnlambdaphi}) 
with $\mathring{\partial}$, using that it commutes past $p$ to yield $\partial$, and 
then applying the Leibniz rule (the second equation in (\ref{GENJacobi})), one computes $ \mathring{\partial} p_2(\Lambda, \Phi)
  - p_2(\partial\Lambda, \Phi) $, using (\ref{p2unpolarized}) for the second term, to confirm that 
  this equals the left-hand side of (\ref{COVariancecond}). 
  This completes the proof of gauge covariance to second order.

After this digression,  
we will show that, in our context, the original field equations (the Maurer-Cartan equations 
for $\Phi$ governed by the $L_{\infty}$ algebra on ${X}$) are equivalent to the field equations 
for $\widehat{\Phi}$ governed by the Maurer-Cartan equations of the $L_{\infty}$ algebra on $\mathring{X}$. 
Thus, as to be expected, one does not loose information by 
passing over to gauge invariant field variables. 
To be more precise, this holds provided the homotopy map in degree $1$ vanishes. However, even without that assumption 
it is always true that the Maurer equations on $X$ imply the Maurer-Cartan equations for a homotopy transferred algebra on 
$\mathring{X}$, 
but the converse is generally not true. 

To see this, recall that the Maurer-Cartan equations can be written as $\pi_1D\exp(\Phi)=0$ and that by 
(\ref{MCEquivalence}) this is equivalent to $D\exp(\Phi)=0$. Thus, 
assuming the original Maurer-Cartan equations 
hold for $\Phi$, we  have 
 \be
  \mathring{D}\exp(\widehat{\Phi}) 
  =\mathring{D}P \exp({\Phi}) 
  = PD\exp({\Phi}) = 0 \;, 
 \ee
where we used (\ref{chainMapCOND}) and (\ref{expphihat}). 
This implies the Maurer-Cartan equations  $\pi_1\mathring{D}\exp(\widehat{\Phi}) =0$ 
for $\widehat{\Phi}$ with respect to 
the homotopy transported $L_{\infty}$ algebra on $\mathring{X}$. 

Next, in order to show the converse (that the Maurer-Cartan equations on $\mathring{X}$ imply the Maurer-Cartan equations on $X$)  
we have to use the homotopy relation lifted to the symmetric algebra, ${\rm id}_{S(X)}=IP+D \frak{H}+\frak{H}D$, 
see (\ref{liftedhomotopy}). We then have 
 \be
 \begin{split} 
 D\exp(\Phi) &= \big(IP+D \frak{H}+\frak{H}D\big) D\exp(\Phi) 
 =IPD\exp(\Phi)\\
 &= I\mathring{D}P\exp(\Phi)= I\mathring{D}\exp(\widehat{\Phi})
  \;, 
 \end{split} 
 \ee
where we used  $D^2=0$ and $\frak{h}_1=0$, so that the lift $\frak{H}$ is trivial on the degree one object $D\exp(\Phi)$, 
and we used (\ref{chainMapCOND}) together with (\ref{expphihat}). 
Thus, if the Maurer-Cartan equation holds on $\mathring{X}$, i.e., if  $\mathring{D}\exp(\widehat{\Phi})=0$, 
we also have $ D\exp(\Phi)=0$, which in turn is equivalent to  the Maurer-Cartan equations on $X$. 
Summarizing, if the degree one homotopy is zero, the Maurer-Cartan equations on $X$ hold if and only 
they hold on $\mathring{X}$, hence the respective equations of motion are equivalent.

Assuming the existence of a cyclic structure or an inner product, the action that yields 
the Maurer-Cartan equations for $\widehat{\Phi}$ is defined in terms of the  homotopy transported  brackets
as 
 \be\label{donwstairsAction} 
  {S}_{\rm KK}[\widehat{\Phi}]
  =\frac{1}{2}\langle \widehat{\Phi}, 
  \mathring{b}_1(\widehat{\Phi})\rangle 
  +\frac{1}{3!} \langle \widehat{\Phi}, 
  \mathring{b}_2(\widehat{\Phi},\widehat{\Phi})\rangle 
  +\frac{1}{4!} \langle \widehat{\Phi}, 
\mathring{b}_3(\widehat{\Phi},\widehat{\Phi},\widehat{\Phi})\rangle +\cdots \;, 
 \ee
where the inner product is the transported one defined in (\ref{transportedcyclci}).  
This action is thus equivalent to the original action $S[\Phi]$: both yield equivalent field equations. 

The above argument relies on $\frak{h}_1=0$, and in this case the transported brackets on fields, and hence the action, 
actually simplify further. Indeed, evaluating the transported 3-bracket (\ref{transport3Bracket}) on $\widehat{\Phi}$ we have 
 \be
  \mathring{b}_3\big(\widehat{\Phi},\widehat{\Phi},\widehat{\Phi}\big) = p\Big\{ b_3\big(\widehat{\Phi},\widehat{\Phi},\widehat{\Phi}\big)
  -3\,b_2\big(\frak{h}_1\big(b_2\big(\widehat{\Phi},\widehat{\Phi}\big)\big), \widehat{\Phi}\big)\Big\} 
  = p\Big\{ b_3\big(\widehat{\Phi},\widehat{\Phi},\widehat{\Phi}\big)\Big\} \;, 
 \ee
where we used that the inclusion on fields is trivial, so that $\iota(\widehat{\Phi})= \widehat{\Phi}$, and 
that $\frak{h}_1=0$. Similarly, all higher $\mathring{b}_n$ just consist of the projection of the original $b_n$. 
Finally, the projector is automatically implemented when `integrating against the constrained $\widehat{\Phi}$\,', 
as in our case follows from (\ref{againstcontrainedaergument}). Therefore, (\ref{donwstairsAction}) reduces to 
the original action but evaluated on $\widehat{\Phi}$: 
 \be
   {S}[\widehat{\Phi}]
  =\frac{1}{2}\langle \widehat{\Phi}, 
  {b}_1(\widehat{\Phi})\rangle 
  +\frac{1}{3!} \langle \widehat{\Phi}, 
  {b}_2(\widehat{\Phi},\widehat{\Phi})\rangle 
  +\frac{1}{4!} \langle \widehat{\Phi}, {b}_3(\widehat{\Phi},\widehat{\Phi},\widehat{\Phi})\rangle +\cdots \;, 
 \ee
where the inner product is now the original one on $X$, and we again suppressed the inclusion using  $\iota(\widehat{\Phi})= \widehat{\Phi}$. 
 This proves the claim in the introduction.

\subsection{Non-linear theory of Kaluza-Klein modes}

We now analyze some details for Kaluza-Klein theory following from the homotopy transfer to gauge invariant fields. 
The full gauge transformations following from (\ref{GaugeFull}) for the fluctuation field $h_{MN}$ are given by 
 \be
  \delta h_{MN} = \partial_M\xi_N+\partial_N\xi_M + {\cal L}_{\xi}h_{MN}\;, 
 \ee
where 
 \be\label{Liexionh} 
{\cal L}_{\xi}h_{MN} = 
\xi^K\partial_K h_{MN}
+\partial_M\xi^K h_{KN} 
+\partial_N\xi^K h_{KM} \;. 
 \ee
Comparing with (\ref{nonlingauge23}) one infers that the only non-trivial $b_n$ for which one argument is a gauge parameter $\Lambda=(\xi_{\mu}, \lambda_m)$
and the other fields 
are $b_1$ and $b_2$, where $b_1$ is defined by the first equation in (\ref{b1upstairsss}),  and  $b_2$ 
is given by the Kaluza-Klein split of (\ref{Liexionh}):
 \be\label{b2onegaugeonefield} 
  \begin{split}
    b_2(\Lambda,h)_{\mu\nu} &= ({\cal L}_{\xi} + {\cal L}_{\lambda}) 
    h_{\mu\nu}+2\partial_{(\mu}\lambda^m\,a_{\nu)m} \;, \\
     b_2(\Lambda,h)_{\mu m} &= ({\cal L}_{\xi} 
     + {\cal L}_{\lambda})a_{\mu m} + \partial_{\mu}\lambda^n\,\varphi_{mn} 
     + \partial_m\xi^{\nu}\, h_{\mu\nu} \;, \\
      b_2(\Lambda,h)_{mn} &= ({\cal L}_{\xi} 
     + {\cal L}_{\lambda})\varphi_{mn} 
     +2\partial_{(m} \xi^{\nu}\, a_{\nu\, n)}\;, 
  \end{split}
 \ee
where the notation indicates purely internal and external Lie derivatives, respectively, i.e., 
 \be
 \begin{split}
  {\cal L}_{\lambda}h_{\mu\nu} &= \lambda^m\partial_m h_{\mu\nu}\;, \\
  {\cal L}_{\lambda}a_{\mu m} &= \lambda^n\partial_n 
  a_{\mu m} + \partial_m\lambda^n a_{\mu n}\;, \\
  {\cal L}_{\lambda}\varphi_{mn} &=\lambda^k\partial_k\varphi_{mn}+2\partial_{(m}\lambda^k \varphi_{n)k}
\;, \end{split}  
 \ee
and 
 \be
 \begin{split}
   {\cal L}_{\xi}h_{\mu\nu} &= \xi^{\rho}\partial_{\rho}h_{\mu\nu}+2\partial_{(\mu}\xi^{\rho}h_{\nu)\rho}\;, \\
   {\cal L}_{\xi} a_{\mu m} &= \xi^{\rho}\partial_{\rho}a_{\mu m}+\partial_{\mu}\xi^{\rho} a_{\rho m}\;, \\
   {\cal L}_{\xi}\varphi_{mn} &= \xi^{\rho}\partial_{\rho}\varphi_{mn} \;. 
 \end{split}
 \ee
All $b_n$ for $n\geq 3$ with one gauge parameter 
and $n-1$ fields are zero.

The above presentation of the gauge transformations is not the familiar one  for the zero-modes, as for instance 
the spin-2 field $h_{\mu\nu}$ transforms under the gauge parameter $\lambda_m$ associated to $a_{\mu m}$. 
Indeed, usually one chooses a particular non-linear Kaluza-Klein ansatz parameterizing the $D$-dimensional fields in terms 
of the lower-dimensional fields 
so that for the zero modes one obtains familiar $n$-dimensional diffeomorphisms w.r.t.~$\xi^{\mu}$ 
and $U(1)^d$ gauge transformations w.r.t.~$\lambda^m$. However,  the gauge transformations including non-zero 
modes then become very non-linear, switching on $b_n$ for $n\geq 3$ (c.f.~eq.~(3.9) in \cite{Baguet:2015xha}, which contains the inverse scalar matrix that induces terms of arbitrary order in fields). 
For this reason, we stick for now  to the above form of the gauge transformations that have the advantage 
of requiring only a $b_2$.

Next, we turn to the homotopy transported $L_{\infty}$ brackets on $\mathring{X}$. 
The differential $b_1=\partial$ in (\ref{b1upstairsss}) is  transported to $\mathring{b}_1=\mathring{\partial}$ in (\ref{mathringpartial}). 
Furthermore, from (\ref{transported2Bracket}) we obtain the transported 2-brackets for one gauge parameter and one field:
 \be
  \mathring{b}_{2}(\mathring{\Lambda}, 
  \mathring{\Phi}) 
  =p(b_2(\mathring{\Lambda},\mathring{\Phi}))\;, 
 \ee
using that the inclusions are trivial in this case. 
Thus, we first have to work out the 2-brackets of two arguments from the restricted space $\mathring{X}$, which are 
 \be\label{b2oncircled}
  \begin{split}
{b}_2(\mathring{\Lambda},\mathring{h})_{\mu\nu} &= {\cal L}_{\mathring{\xi}}\mathring{h}_{\mu\nu} + \mathring{\lambda}^m\partial_m 
    \mathring{h}_{\mu\nu}+2\partial_{(\mu}
    \mathring{\lambda}^m\,
    \mathring{a}_{\nu)m}\;,  \\
     {b}_2(\mathring{\Lambda},\mathring{h})_{\mu m} &= {\cal L}_{\mathring{\xi}} \mathring{a}_{\mu m}
     +{\mathring{\lambda}}^n\partial_n\mathring{a}_{\mu m} + \partial_{\mu}\mathring{\lambda}^n\,
     \mathring{\varphi}_{mn}\;,  \\
      b_2(\mathring{\Lambda},\mathring{h})_{mn} &= {\cal L}_{\mathring{\xi}} \mathring{\varphi}_{mn} 
     + {\mathring{\lambda}}^k\partial_k
     \mathring{\varphi}_{mn} \;, 
  \end{split}
 \ee
and then project down to $\mathring{X}$. However, 
using that in the latter space the parameters only carry zero modes one sees 
that the result already obeys all divergence constraints. 
Therefore, no explicit projection is required, and we can immediately write 
 \be\label{b2curcleddd} 
  \begin{split}
\mathring{b}_2(\mathring{\Lambda},\mathring{h})_{\mu\nu} &= {\cal L}_{\mathring{\xi}}\mathring{h}_{\mu\nu} + \mathring{\lambda}^m\partial_m 
    \mathring{h}_{\mu\nu}+2\partial_{(\mu}
    \mathring{\lambda}^m\,
    \mathring{a}_{\nu)m} \;, \\
     \mathring{b}_2(\mathring{\Lambda},\mathring{h})_{\mu m} &= {\cal L}_{\mathring{\xi}} \mathring{a}_{\mu m}
     +{\mathring{\lambda}}^n\partial_n\mathring{a}_{\mu m} + \partial_{\mu}\mathring{\lambda}^n\,
     \mathring{\varphi}_{mn} \;, \\
      \mathring{b}_2(\mathring{\Lambda},\mathring{h})_{mn} &= {\cal L}_{\mathring{\xi}} \mathring{\varphi}_{mn} 
     + {\mathring{\lambda}}^k\partial_k
     \mathring{\varphi}_{mn} \;. 
  \end{split}
 \ee 
Next we can compute  the transported 3-bracket with (\ref{transport3Bracket}), which  is actually zero, 
  \be 
\mathring{b}_3(\mathring{\Lambda},
\mathring{\Phi},\mathring{\Phi}) = 
-2pb_2\big(\frak{h}(b_2(\mathring{\Lambda},
\mathring{\Phi})),\mathring{\Phi}\big) = 0 \;. 
 \ee
This follows since   the relevant homotopy map 
(\ref{homotopy0Torus3}) in degree zero only takes internal divergencies of the Kaluza-Klein scalar and vector  components, but all 
internal divergencies of $b_2(\mathring{\Lambda},\mathring{\Phi})$ in (\ref{b2oncircled}) vanish due to the first (gauge parameter) argument having only zero modes. Thus, there is no $b_3$ in the $L_{\infty}$ algebra on $\mathring{X}$ that would contribute  to the gauge transformations. 
Similarly, there are no higher $\mathring{b}_n$ for one gauge parameter and $n-1$ fields. 
Therefore, the gauge transformations (\ref{nonlingauge237}) reduce to 
   \be\label{nonlingauge23237}
   \delta_{\Lambda}\widehat{\Phi} = \mathring{b}_1\big(\widehat{\Lambda}\big) 
   + \mathring{b}_2\big(\widehat{\Lambda}, \widehat{\Phi}\big)  \;. 
  \ee
  
Dropping  the circle and hat for the remainder of this section, we have 
with (\ref{b2curcleddd}) 
the full gauge transformations 
 \be\label{semifinalgauge} 
  \begin{split}
   \delta h_{\mu\nu} &= 2\partial_{(\mu}\xi_{\nu)} + ({\cal L}_{\xi}+\lambda^k\partial_k)h_{\mu\nu} 
   +2\partial_{(\mu} {\lambda}^m \,{a}_{\nu)m} \;, \\
   \delta a_{\mu m} &= \partial_{\mu}\lambda_m + ({\cal L}_{\xi}+\lambda^k\partial_k)a_{\mu m} + \partial_{\mu}\lambda^n \varphi_{mn} \;, \\
   \delta \varphi_{mn} &= ({\cal L}_{\xi}+\lambda^k\partial_k) \varphi_{mn}\;. 
  \end{split} 
 \ee
Recall that $a_{\mu m}$ and $ \varphi_{mn}$, as components of the $\widehat{\Phi}$, are subject to the 
constraint that all internal divergencies are zero, and that the above gauge transformations preserve these constraints 
because  the gauge parameters carry only zero modes. 

While simple, the above gauge transformations are still non-standard for the zero modes of $h_{\mu\nu}$ and $a_{\mu m}$, 
due to the last terms in the first and second line of (\ref{semifinalgauge}). 
These terms, however, can be removed by field redefinitions. 
To this end, it is convenient to use matrix notation, 
viewing  $\varphi=(\varphi_{mn})$ as a $d\times d$ matrix and 
$a_{\mu} = ( a_{\mu m})$ as a (dual) $d$-vector and $\lambda = (\lambda^m)$ as  a $d$-vector.
 We also write the internal background 
metric as $\bar{G}=(\bar{G}_{mn})$. 
Then the  gauge transformations 
can be written as 
 \be\label{indexfreegauge} 
 \begin{split} 
  \delta h_{\mu\nu} &= 2\partial_{(\mu}\xi_{\nu)} +({\cal L}_{\xi}+\lambda\cdot\partial)h_{\mu\nu} +2\partial_{(\mu} \lambda \cdot a_{\nu)} \;, \\
  \delta a_{\mu} &= (\bar{G}+\varphi) \partial_{\mu} \lambda + ({\cal L}_{\xi}+\lambda\cdot\partial)a_{\mu} \;, \\
  \delta \varphi &= ({\cal L}_{\xi}+\lambda\cdot\partial)\varphi \;, 
 \end{split} 
 \ee
where we used the short-hand notation $\lambda\cdot\partial = \lambda^m\partial_m$. 
This suggests the definition of a background independent internal metric 
 \be
  G_{mn} := \bar{G}_{mn}  +\varphi_{mn}  \qquad \Leftrightarrow \qquad G = \bar{G}+\varphi \;, 
 \ee
and a new vector field 
 \be
 A_{\mu}:= G^{-1}a_{\mu}  \qquad \Leftrightarrow \qquad A_{\mu}{}^m =G^{mn} \,a_{\mu n}\;. 
 \ee
Its gauge transformation then follows with (\ref{indexfreegauge}):
 \be
 \begin{split} 
  \delta A_{\mu}  &= \delta \big[ G^{-1}\big] a_{\mu} + G^{-1} \delta a_{\mu} \\
 &= ({\cal L}_{\xi}+\lambda\cdot\partial) \big[ G^{-1}\big] a_{\mu}
 + G^{-1}\big(G \partial_{\mu} \lambda + ({\cal L}_{\xi}+\lambda\cdot\partial)a_{\mu}\big) \\
 &=\partial_{\mu}\lambda +({\cal L}_{\xi}+\lambda\cdot\partial) A_{\mu}\;. 
 \end{split} 
 \ee
This is the desired covariant gauge transformation. Similarly, redefining the spin-2 field as 
 \be
  h_{\mu\nu}' := h_{\mu\nu} -a_{\mu}\,{G}^{-1} \,a_{\nu}  \qquad \Leftrightarrow \qquad
   h_{\mu\nu}' = h_{\mu\nu} -a_{\mu m}G^{mn}a_{\nu n}\;, 
 \ee 
 one finds by a similar computation that 
  \be
    \delta h'_{\mu\nu} = 2\partial_{(\mu}\xi_{\nu)} +({\cal L}_{\xi}+\lambda\cdot\partial)h'_{\mu\nu} \;, 
  \ee 
which is the expected covariant transformation. More precisely, these are standard diffeomorphisms and $U(1)^d$ gauge transformation 
on zero modes, while on non-zero modes the  $U(1)^d$ gauge parameter $\lambda^m$ acts via an extra transport term.

The gauge transformations are now such that they can be rewritten in a background independent form for 
 \be
  \begin{split}
   g_{\mu\nu} := \eta_{\mu\nu}+h'_{\mu\nu}\;, \qquad 
   A_{\mu}{}^{m}  \;, \qquad 
   G_{mn} := \delta_{mn} + \varphi_{mn} \;, 
  \end{split} 
 \ee
as  
 \be\label{finalnewgauge} 
  \begin{split} 
   \delta g_{\mu\nu} &= {\cal L}_{\xi}g_{\mu\nu} +\lambda^m\partial_m g_{\mu\nu}\;, \\
   \delta A_{\mu}{}^{m} &= \partial_{\mu}\lambda^m +\lambda^n \partial_n A_{\mu}{}^{m} + {\cal L}_{\xi}A_{\mu}{}^{m} \;, \\
   \delta G_{mn} &= {\cal L}_{\xi}G_{mn}  +\lambda^k\partial_k G_{mn}\;. 
  \end{split}
 \ee 
Thus, all fields transform in the standard way under $n$-dimensional diffeomorphisms,  while the transport term $\lambda^k\partial_k$ 
encodes the $U(1)^d$ action on the massive modes.\footnote{One is naturally  tempted to perform the usual parameter redefinition 
$\lambda^m\rightarrow \lambda^m+\xi^{\mu}A_{\mu}{}^{m}$, but this would be illegal as $A_{\mu}{}^{m}$ carries non-zero modes, 
while $\lambda^m$ does not.} 
Truncating to the zero modes, the above simply encodes the  standard Kaluza-Klein ansatz that embeds the lower-dimensional fields into 
the $D$-dimensional metric. Indeed, rewriting the original expansion (\ref{backgroundmetricexp}) of the metric, which we took to be exact, in terms of the new fields we have 
 \be\label{zeromodeaction} 
 \begin{split} 
  G_{MN} = \begin{pmatrix} G_{\mu\nu} &  G_{\mu n}  \\
 G_{m \nu}  & G_{mn} \end{pmatrix} 
 &=  \begin{pmatrix} \eta_{\mu\nu} + h_{\mu\nu} &  a_{\mu n}  \\
 a_{\nu m }  & \delta_{mn}+\varphi_{mn}  \end{pmatrix} \\
 &=  \begin{pmatrix} g_{\mu\nu} + G_{mn}A_{\mu}{}^{m} A_{\nu}{}^{n}  
 &  G_{mn} A_{\mu}{}^{n} \\
  G_{mn}A_{\nu}{}^{n} & G_{mn} \end{pmatrix} \;, 
 \end{split} 
 \ee
which is the standard Kaluza-Klein form (without a Weyl rescaling $g_{\mu\nu}\rightarrow (\det G)^{-\frac{1}{n-2}}g_{\mu\nu}$ that brings the 
lower-dimensional metric to Einstein frame but is not needed for canonical gauge transformations). 
One should recall, however, that here the fields must be thought of as expanded around the constant background, 
with the fluctuations being the gauge invariant  redefinitions of the original fluctuations, 
obeying the appropriate constraints.

\section{Conclusion and Outlook}

In this paper we have shown, in the context of Kaluza-Klein theories that keep all higher Kaluza-Klein modes, 
how to define, to any order in perturbation theory, new field variables that are gauge invariant under all 
non-zero mode gauge transformations, which are spontaneously broken. 
This yields an explicit understanding of the Higgs mechanism 
that renders the spin-2, vector and higher tensor modes massive. To this end we used  the framework of homotopy algebras, 
such as the $L_{\infty}$ algebras that encode field theories, together with the notion of homotopy transfer. 
In this case, the homotopy transfer in particular maps the space of fields subject to an infinite-dimensional 
gauge symmetry to a space of fields that are \textit{invariant} under all non-zero mode gauge transformations, 
while transforming covariantly under the remaining finite-dimensional zero-mode gauge transformations. 
Assuming only that the gauge algebra of the theory closes off-shell (equivalently, that the $L_{\infty}$ 
algebra of the gauge subsector exists independently) the construction of gauge invariant field variables 
is off-shell: no equations of motion need to be solved in order to establish gauge invariance. 
This fact in turn implies a surprising result: the action for the gauge invariant fields is 
obtained from the original action by simply replacing the original fields by the gauge invariant ones. 
As the latter are subject to constraints, certain couplings will disappear in the final action. 
Operationally, going to gauge invariant variables is thus equivalent to fixing a gauge, with the constraints 
that the gauge invariant variables satisfy being reinterpreted as gauge fixing conditions, but it is 
important to recall that the action is valid for perturbations \textit{in any gauge}. 

The results of this paper are just the first step in a larger program that aims to systematize, and enable 
for so far inaccessible backgrounds, 
the computation of boundary correlation functions from a bulk Kaluza-Klein theory with AdS factor. 
After eliminating the infinite-dimensional gauge redundancy associated to the non-zero mode gauge parameters, 
the next step is to solve the bulk equations so as to determine the on-shell action that, according to 
the AdS/CFT correspondence, computes the boundary correlation functions. 
Remarkably, also this step has an interpretation as homotopy transfer, as was established in 
\cite{Bonezzi:2025bgv} for scalar and Yang-Mills theories. 
In a sequel to this paper we will generalize the homotopy transfer interpretation for the massive Kaluza-Klein modes 
of the torus displayed here to a large class of backgrounds \cite{Eloy:2025xyz}. This will  provide the means to then extend, for these geometries,
the homotopy transfer to the boundary established in \cite{Bonezzi:2025bgv}, which we expect to implement the Witten diagrams.

 \section*{Acknowledgments} 

We thank Roberto Bonezzi and especially Christoph Chiaffrino for helpful discussions. 

\noindent
This work is funded by the Deutsche Forschungsgemeinschaft (DFG, German Research Foundation), ``Rethinking Quantum Field Theory", Projektnummer 417533893/GRK2575.

\appendix

\section{St\"uckelberg formulation of Proca via homotopy transfer}

As a simple toy model, in this appendix we explain  the definition of gauge invariant variables 
for the St\"uckelberg formulation of Proca theory. The latter gives a gauge invariant formulation 
of massive spin-1 in terms of  the Lagrangian  
 \be
  {\cal L} (A_{\mu}, \varphi) = -\frac{1}{4} F^{\mu\nu}F_{\mu\nu} -\frac{1}{2} D^{\mu}\varphi D_{\mu}\varphi \;, 
 \ee
where, of course, $F_{\mu\nu}=\partial_{\mu}A_{\nu}-\partial_{\nu}A_{\mu}$, and 
 \be
  D_{\mu}\varphi = \partial_{\mu}\varphi - m A_{\mu} \;. 
 \ee
The action is gauge invariant under 
 \be
  \delta A_{\mu} = \partial_{\mu}\lambda \;, \qquad \delta\varphi = m\lambda\;. 
 \ee
The field equations are 
 \be
  \begin{split}
   \partial_{\mu} F^{\mu\nu} + mD^{\nu}\varphi &=0 \;, \\
   \partial^{\mu}D_{\mu}\varphi &= 0\;. 
  \end{split} 
 \ee 
It is customary to pick  unitary gauge $\varphi=0$, in which case the second equation reduces to 
the subsidiary condition $\partial^{\mu}A_{\mu}=0$, but in the following we rather go to gauge invariant variables
via homotopy transfer.

Encoding this theory in a four term chain complex with degrees from $-1$ to $2$, with fields ${\cal A}=(A_{\mu},\varphi)$ in degree 0 and 
field equations ${\cal E}=(E^{\mu}, E)$ in degree one,  we have the chain complex 
 \begin{figure}[h]
	\begin{centering}
		\label{fig:homotopy transfer2}
		\begin{tikzpicture}
			\node at (3,0) {$X_{-1}$}; 
			\draw[->] (3.5,0)--(4.5,0);
			\node at (4,0.3) {$\partial$};
			\node at (5,0) {$X_0$}; 
			\draw[->] (5.5,0)--(6.5,0);
			\node at (6,0.3) {$\partial$};
			\node at (7,0) {$X_1$}; 
			\draw[->] (7.5,0)--(8.5,0);
			\node at (8,0.3) {$\partial$};
			\node at (9,0) {$X_2$}; 
			\draw[->] (4.9,0.3) to [bend right=45] (3,0.3);
			\node at (4,1) {$\frak{h}_0$};
			\draw[->] (6.9,0.3) to [bend right=45] (5.1,0.3);
			\node at (6,1) {$0$};
			\draw[->] (9,0.3) to [bend right=45] (7.1,0.3);
			\node at (8,1) {$\frak{h}_2$};
		\end{tikzpicture}
		\caption{Proca chain complex}
	\end{centering}
\end{figure}
and the differentials:
 \be
  \begin{split}
   \partial_{-1}(\lambda)&= \begin{pmatrix} \partial_{\mu}\lambda \\ m\lambda \end{pmatrix}\;,  \\
   \partial_0({\cal A}) &= \begin{pmatrix} \square A^{\nu} -\partial^{\nu}(\partial\cdot A)  +m\partial^{\nu}\varphi -m^2A^{\nu} \\ \square \varphi -m\partial^{\mu}A_{\mu} \end{pmatrix}\;,  \\
   \partial_1({\cal E}) &= \partial_{\mu}E^{\mu} -mE\;. 
  \end{split}
 \ee 
 
 This should be homotopy equivalent to the two-term complex for strict Proca, where projection and inclusion are only non-trivial in degrees 
 zero and one:
  \be
   \begin{split}
    p_0({\cal A}) &
    = A_{\mu} - \frac{1}{m}\partial_{\mu}\varphi \;, \qquad
    \iota_0(A_{\mu}) = \begin{pmatrix} A_{\mu} \\ 0 \end{pmatrix}\;,  \\
    p_1({\cal E}) &= E^{\mu} \;, \qquad\qquad \qquad   \iota_1(E^{\mu}) = \begin{pmatrix} E^{\mu} \\ \frac{1}{m}\partial_{\nu}E^{\nu}
     \end{pmatrix}\;. 
   \end{split} 
  \ee
It is straightforward to verify the chain map conditions $\bar{\partial}\circ p = p \circ \partial$ 
and $\partial\circ \iota = \iota\circ \bar{\partial}$. In particular, this requires the non-trivial inclusion $\iota_1$. 
We next give three homotopy maps, $\frak{h}_0$, $\frak{h}_1$, and $\frak{h}_{2}$ so that 
 ${\rm id} - \iota p  = \partial \frak{h} + \frak{h} \partial$: 
  \be
    h_0({\cal A}) = h_0\begin{pmatrix}A_{\mu}\\ \varphi \end{pmatrix} = \frac{1}{m}\varphi \;,  \qquad 
    h_1({\cal E})=0\;, \qquad
    h_2({\cal N}) = -\frac{1}{m} \begin{pmatrix} 0 \\ {\cal N} \end{pmatrix}  \;. 
 \ee 
 It is straightforward to verify the homotopy relations.

\bibliographystyle{utphys}
\bibliography{refs}

\end{document}